\documentclass[12pt,a4paper]{article}
\usepackage[utf8]{inputenc}
\usepackage{amsmath}
\usepackage{amssymb}
\usepackage{graphicx}
\usepackage[all]{xy}
\usepackage{float}
\usepackage{amscd}
\usepackage{color}
\usepackage{amsfonts}
\usepackage{amstext}
\usepackage{epstopdf}
\usepackage[T1]{fontenc}
\usepackage{geometry}
\definecolor{myurlcolor}{rgb}{0,0,0.7}
\definecolor{myrefcolor}{rgb}{0.8,0,0}
\usepackage[unicode=true,pdfusetitle, bookmarks=true,bookmarksnumbered=false,bookmarksopen=false, breaklinks=false,pdfborder={0 0 0},backref=false,colorlinks=true, linkcolor=myrefcolor,citecolor=myurlcolor,urlcolor=myurlcolor]{hyperref}

\usepackage{stmaryrd}
\def\NN {{\mathbb N}}     %% natural numbers
\def\PP {{\mathbb P}}     %% projective
\def\RR {{\mathbb R}}     %% real numbers
\def\XX {{\mathbb X}}     %% sphere
\def\lo  {\longmapsto}

\def\lw  {\longrightarrow}
\def\mc {\mathcal}
\def\mk {\mathfrak}

\def\rw  {\rightarrow}

\global\long\global\long\global\long\def\SET#1#2{\mbox{\ensuremath{\ensuremath{\left\lbrace\left. #1\ \right|\ #2  \right\rbrace }}}}

\date{}

\newtheorem{theorem}{Theorem}

\newcommand{\tr}{\mbox{Tr}}
\newcommand{\bra}[1]{\mbox{$\langle #1 |$}}
\newcommand{\ket}[1]{\mbox{$| #1 \rangle$}}
\newcommand{\bk}[2]{\ensuremath{\langle #1 | #2 \rangle}}
\newcommand{\kb}[2]{\ensuremath{| #1 \rangle\!\langle #2 |}}
\newcommand{\Ad}{\mbox{Ad}}
\begin{document}

\title{Multipartite quantum correlations: symplectic and algebraic geometry approach}

\author{Adam Sawicki$^{1}$, Tomasz Maci\k{a}\.zek$^1$, Micha{\l} Oszmaniec$^{1,2}$, \\ Katarzyna Karnas$^1$, Katarzyna Kowalczyk-Murynka$^1$, Marek Ku\'{s}$^1$   
}

\date{}

\maketitle

$^{1}$Center for Theoretical Physics, Polish Academy of Sciences, Al.
Lotnik\'ow 32/46, 02-668 Warszawa, Poland

$^{2}$ICFO-Institut de Ciencies Fotoniques, The Barcelona Institute of
Science and Technology, 08860 Castelldefels (Barcelona), Spain

\begin{abstract}
We review a geometric approach to classification and examination of quantum correlations in composite systems. Since quantum information tasks are usually achieved by manipulating spin and alike systems or, in general, systems with a finite number of energy levels, classification problems are usually treated in frames of linear algebra. We proposed to shift the attention to a geometric description. Treating consistently quantum states as points of a projective space rather than as vectors in a Hilbert space
we were able to apply powerful methods of differential, symplectic and algebraic geometry to attack the problem of equivalence
of states with respect to the strength of correlations, or, in other
words, to classify them from this point of view. Such classifications
are interpreted as identification of states with `the same correlations
properties' i.e. ones that can be used for the same information
purposes, or, from yet another point of view, states that can be
mutually transformed one to another by specific, experimentally
accessible operations. It is clear that the latter characterization
answers the fundamental question `what can be transformed into what
\textit{via} available means?'. Exactly such an interpretation, i.e,
in terms of mutual transformability, can be clearly formulated in terms
of actions of specific groups on the space of states and is the
starting point for the proposed methods.
\end{abstract}

\section{Introduction} \label{sec:wprowadzenie}

Quantum entanglement - a direct consequence of linearity of quantum
mechanics and the superposition principle - is one of the most
intriguing phenomena distinguishing the quantum and classical description
of physical systems. Quantum correlated (e.g., entangled) states of
composite systems possess features unknown in the classical world, like
the seemingly paradoxical nonlocal properties exhibited by the famous
Einstein-Podolsky-Rosen analysis of completeness of the quantum theory.
Recently, with the development of quantum information theory they came
to prominence as the main resource for several applications aiming at
speeding up and making more secure information transfers (see, e.g.,
\cite{Horodeccy}). A novel kind of quantum correlations,
called \textit{quantum discord}, different from entanglement, but also
absent in the macroscopic world, was discovered \cite{OZ01},
\cite{HV01} adding one more element to ``the mysteries of quantum
mechanics'' as seen from the classical point of view.

Although typically a quantum system, as, e.g., a harmonic oscillator or
a hydrogen atom,  is described in terms of an infinite-dimensional
Hilbert space, for most quantum-information applications the
restriction to finite dimensions suffices, since usually the active
role in information processing play only spin degrees of freedom or
only few energy levels are excited during the evolution.

From the mathematical point of view such finite-dimensional quantum
mechanics seems to mount a smaller challenge than in the
infinite-dimensional case - the tool of choice here is linear algebra
rather than functional analysis. Nevertheless, understanding of
correlations in multipartite finite dimensional quantum systems  is
still incomplete, both for systems of distinguishable particles \cite{RevEnt2017} as well as for ones 
consisting of non-distinguishable particles like bosons and 
fermions \cite{RedMat,MKFER2001, CORNOND2002, GKM11}.

The statistical interpretation of quantum mechanics disturbs a bit the
simple linear-algebraic approach to quantum mechanics - vectors corresponding to a state
 (elements of a finite-dimensional Hilbert space $\mathcal{H}$)
should be of unit norm. Obviously, physicist are accustomed to cope
with this problem in a natural way by ``normalizing the vector and
neglecting the global phase''. Nevertheless, it is often convenient to
implement this prescription by adopting a suitable mathematical
structure, the projective space $\mathbb{P}(\mathcal{H})$, already from
the start\footnote{Equivalently, it is possible to incorporate the redundancy of the global phase by identifying pure states with orthogonal projectors onto one-dimensional subspaces of $\mathcal{H}$. However, for the sake of convenience, in this exposition we decided to treat pure states as elements of $\mathbb{P}(\mathcal{H})$}. The projective space is  obtained from the original Hilbert
space $\mathcal{H}$ by identifying vectors\footnote{We will use
exchangeably the Dirac notation, $\ket{\psi}$, etc., and the short one
$\psi$ etc.\ for elements (vectors) of $\mathcal{H}$.} differing by a
scalar, complex, non-zero factor, $\ket{\psi}\cong c\ket{\psi}$. We
will denote elements (points) of $\mathbb{P}(\mathcal{H})$ by $u,v,x$,
etc.\ and, if we want to identify a particular equivalence class of the
vector $\ket{\psi}$, by $[\psi]$, etc. 

Obviously, both approaches, the linear-algebraic (plus normalization
and neglecting the global phase) picture and the projective one are
equivalent. Following the former we loose linearity, so which
advantages we could expect instead? We answered this question in our
paper \cite{SHK11}, where we propose, by working in the projective
space, to apply completely new (in this context) techniques to analyze
the phenomenon of entanglement. The approach has given a deeper insight
into the unexpectedly rich geometric structure of the space of states
and enabled the use of recently developed advanced methods of complex
differential, algebraic and symplectic geometry.

The most efficient characterization of quantum correlations is achieved
by identifying states that are `equally correlated' or, in other words,
states that can be mutually transformed \textit{via} methods allowed by
rules of quantum mechanics without destroying quantum correlations. 
\begin{figure}[ht]\label{fig:corr_pres}
\centering
\resizebox{0.6\linewidth}{!}{\includegraphics{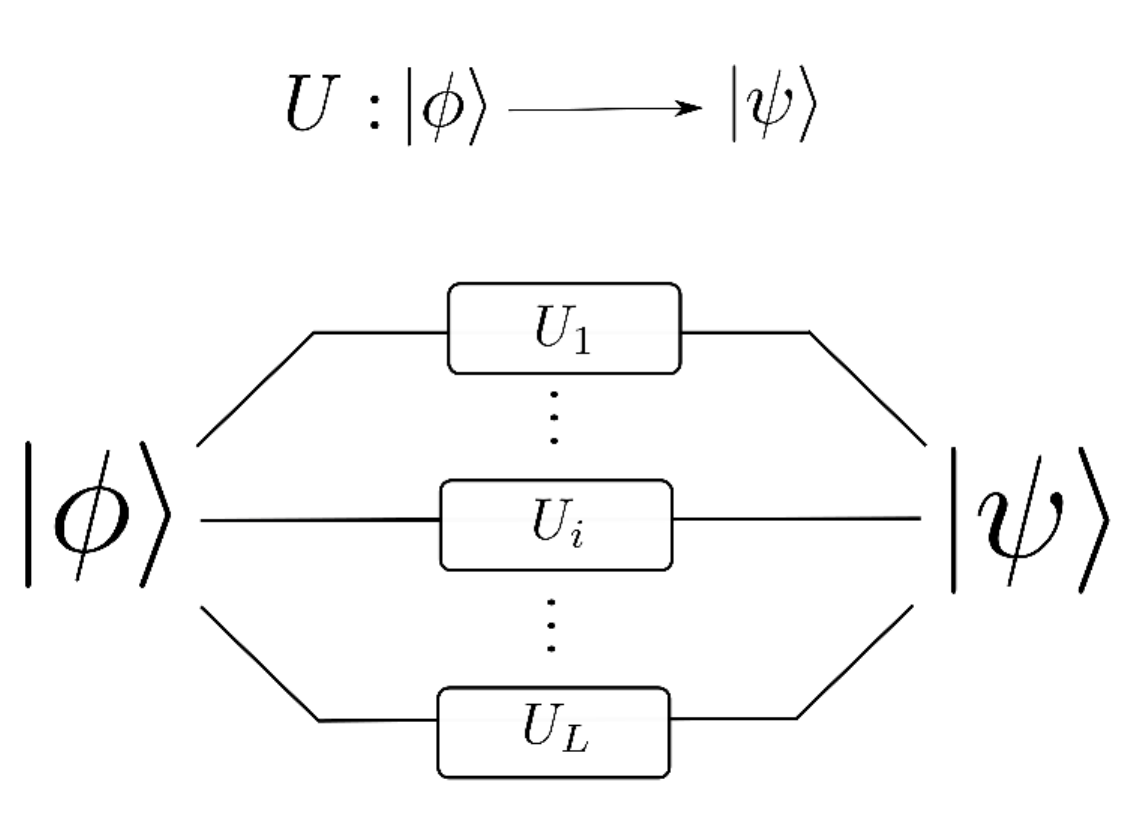}}
\caption{ A correlation-preserving  transformation $U\in SU(N_1)\times\ldots\times SU(N_L)$ acting on the entangled state $|\phi\rangle$ of $L$ particles. Each $U_i$ acts on the one-particle Hilbert space $\mathcal{H}_i$ separately.  }
\end{figure}
From the perspective of quantum information theory it is then natural to consider quantum operations that are \textit{local},
i.e.\ restricted to act independently on each subsystem. As required by
quantum mechanics, such operations are unitary in the relevant Hilbert
space $\mathcal{H}$, but form only a subgroup of the whole group
$U(\mathcal{H})$ as limited by the restriction to subsystems. For these
reasons they are called \textit{local unitary} (LU) operations \cite{Acin2001,Ferstrate2003,Krauss2009,Vrana2010}. A
complementary insight is provided by adding more operations, like
measurements performed locally, i.e., each restricted to a single
subsystem, and possible communication among subsystems \textit{via}
classical means of communication. Such operations are dubbed  LOCC -
\textit{Local Operations with Classical Communication} \cite{Nielsen1999}. They usually
destroy purely quantum correlations, but in any case do not increase
them, so if one state can be transformed to another one \textit{via}
LOCC transformation, the latter is not more (and usually less)
correlated than the former. This allows quantifying correlations by
constructing \textit{entanglement measures}, quantities characterizing
states that remain unchanged under LU operations and do not increase
under LOCC ones.  Technically it is convenient to group together states
related to each other by invertible stochastic local operations and
classical communication (SLOCC). Two states, $\ket{\phi}$ and
$\ket{\psi}$ belong to the same SLOCC class if both transformations,
$\ket{\psi}\rightarrow\ket{\phi}$ and $\ket{\phi}\rightarrow\ket{\psi}$
can be achieved by means of LOCC with non-zero probability \cite{SLOCC1,SLOCC2,SLOCC3}. Hence, in
general,  if two states belong to two distinct SLOCC classes, they have
different correlation properties and might be not exchangeable for
achieving the same quantum informational tasks.

As mentioned above the LU operations form a subgroup $K$ of the whole
unitary group $U(\mathcal{H})$ of the relevant Hilbert space
$\mathcal{H}$ (since the global phase of a state is irrelevant, we can
use the special unitary group $SU(\mathcal{H})$ instead). The SLOCC
also form a group $G$ that happens to be the complexification of $K$,
i.e., $G = K^\mathbb{C}$. In contrast to $K$, which is compact as a
subgroup of the unitary group, $G$ is not compact. The exact forms of
$K$ and, consequently, $G$ depend on the problem in question (number
and dimensionality of subsystems, distinguishability of particles
constituting them). The actions of $K$ and $G$ on $\mathcal{H}$ are
naturally and easily transferred to the projective space 
$\mathbb{P}(\mathcal{H})$  \textit{via} 
\begin{equation}\label{eq:pureACTION}
\Phi_{V}([\psi])=[V\psi]\ ,
\end{equation}  where $V$ belongs either to $K$ or to $G$. In other words, after the action of a matrix from the representation of $K$ or $G$ on a vector from $\mathcal{H}$, one has to projectivise the resulting vector to $\mathbb{P}(\mathcal{H})$. Projectivisation means setting the norm to $1$ by rescaling the vector by the inverse of its norm and neglecting the global phase.

The projective space has a reach geometrical structure. In particular,
in a natural way, it is a K\"ahler manifold, i.e., a complex symplectic
Riemannian manifold on which all three structures are compatible.
Symplecticity means that it can be treated as a kind of a classical
phase space for a Hamiltonian system. The action of a group on such a
manifold reflects symmetries of the dynamical system, and the methods
of analysis of such systems can be borrowed from classical mechanics.
In modern formulations of classical mechanics (see, e.g.,
\cite{arnold13}, \cite{guillemin84}) one employs the full power of
symplectic geometry. See Appendix for some background information concerning the action of groups on symplectic manifolds.

The geometric treatment of correlations outlined above was
restricted only to pure states. There exists, however, a natural extension
to mixed states. Instead of the projective space
$\mathbb{P}(\mathcal{H})$ one has to consider the set of all density matrices 
\begin{equation}
\mathcal{D}\left( \mathcal{H}\right) =\SET{\rho}{\rho\in \mathrm{End}(\mathcal{H}),\ \rho\geq 0,\ \mathrm{tr}(\rho)=1 }\ .
\end{equation}
This set can be decomposed into disjoint union of manifolds of isospectral density matrices.  The manifold of isospectral density matrices $\mathcal{O}_\rho$ is an orbit of
the whole $SU(\mathcal{H})$ group through a chosen mixed state $\rho$ (i.e.,
the set of all density matrices obtained from the chosen one
$\rho$ by unitary conjugations\footnote{In other words, a set of isospectral density matrices consists of states that have the same (ordered) spectrum as a given referential state.}, $U\rho U^\dagger$, $U\in
SU(\mathcal{H})$). Such an orbit possesses a natural K\"ahlerian structure given by
the so-called Kirillov-Kostant-Souriau form. This observation allows to
apply the same geometric techniques as in the pure-state case. In this work we focus only on actions of compact groups on $\mathcal{O}_\rho$ which are given by unitary conjugations,
\begin{equation}\label{eq:mixedSTATES}
\Phi_{V}(\sigma) = V \sigma V^\dagger\ ,
\end{equation}
where $\sigma\in \mathcal{O}_\rho$ and  $V\in K$.

\section{Symplectic structures in spaces of states}\label{sec:sympstates}

As outlined in previous section, one of the basic problems in the
theory of quantum correlations is the classification of states with
respect to local operations performed independently on subsystems of a
given system by two classes of such operations LU and SLOCC, both being
groups acting on the space of states, a unitary one $K$ and its
complexification  $G = K^\mathbb{C}$, respectively. The space of pure
states is the projective space $\mathbb{P}(\mathcal{H})$, and for mixed
states, the space of isospectral (i.e., possessing the same eigenvalues)
density matrices. For a composite system of distinguishable particles the underlying Hilbert space  is the tensor product of the Hilbert spaces of the subsystems.  In the case of indistinguishable particles the Hilbert space is an appropriate (anti)-symmetrization of this product.

In both cases of pure and mixed states the state spaces described above have a natural structure of symplectic manifolds (see Appendix for a brief description of relevant mathematical concepts used in the paper). 

For applications in the theory of quantum correlations we will consider the following particular instances.

\begin{enumerate}
\item $M=\mathbb{P}(\mathcal{H})$ (the complex projective space),
    and
\begin{enumerate}
\item 
	for pure states of $L$ distinguishable particles  the 	Hilbert space is the tensor product, $\mc H=\mc H_1\otimes \mc H_2\otimes\ldots\otimes\mc H_L$, where $\mc H_i$ is the $N_i$-dimensional Hilbert space of $i$-th
    particle, the action of $K=SU(N_1)\times\ldots\times SU(N_L)$ on $\mc H$ is defined in the natural way in terms of the tensor product, i.e. $U_1\otimes\cdots\otimes U_L\cdot (\psi_1\otimes\cdots\otimes\psi_L)=U_1\psi_1\otimes\cdots\otimes U_L\psi_L$,    the complexified group is thus $G=K^\mathbb{C}=SL(N_1,\mathbb{C})\times \ldots\times SL(N_L,\mathbb{C})$, and the Lie algebras read
    $\mathfrak{k}=\mathfrak{su}(N_1)\oplus\ldots
    \oplus\mathfrak{su}(N_L)$,
    $\mathfrak{g}=\mathfrak{sl}(N_1,\mathbb{C})\oplus\ldots
    \oplus\mathfrak{sl}(N_L,\mathbb{C})$;

\item for pure states of $L$ $d$-state bosons
    $\mathcal{H}=S^L\mc H_1$ (the $L$-th symmetrized tensor
    power of $\mc H_1$), where $\mc H_1$ is the Hilbert space
    of a single boson, the group $K=SU(d)$ acts diagonally, $U\cdot(\psi_1\vee\cdots\vee\psi_L)=U\psi_1\vee\cdots\vee\ U\psi_L$, and, consequently, $G=K^\mathbb{C}=SL(d,\mathbb{C})$,  $\mathfrak{k}=\mathfrak{su}(d)$,
    $\mathfrak{g}=\mathfrak{sl}(d,\mathbb{C})$;

\item for pure states of $L$ $d$-state fermions,
    $\mathcal{H}=\bigwedge^L\mc H_1$ (the $L$-th anti-symmetrized
    tensor power of $\mc H_1$), where $\mc H_1$ is the Hilbert
    space of a single fermion, the group $K=SU(d)$ again acts diagonally (i.e, by the same element of the group on each factor in the anti-symmetrized product), $G=K^\mathbb{C}=SL(d,\mathbb{C})$,
   $\mathfrak{k}=\mathfrak{su}(d)$,
    $\mathfrak{g}=\mathfrak{sl}(d,\mathbb{C})$.
\end{enumerate}

\item Isospectral density matrices  $\mc O_\rho$  for systems of
    distinguishable particles, with the Hilbert space and the groups
    $K$  as in 1(a) above, but now acting \textit{via}
    conjugations  on elements of $\mathcal{O}_\rho$\footnote{In fact this action can be related to the adjoint action of $K$ on the Lie algebra
    $\mathfrak{su}(N_1\cdot \ldots \cdot N_L)$. Density matrices are not exactly elements of this Lie algebra since the latter are
    traceless. Nevertheless, a shift by a constant multiple of the
    identity yields the desired normalization to $\tr\rho=1$, and
    the requirement of positivity restricts them to a subset of so
    `shifted' Lie algebra.}  (see Eq.\eqref{eq:mixedSTATES}). 
\end{enumerate}

For $\mathbb{P}(\mathcal{H})$ the Kirillov-Kostant-Souriau form providing a symplectic structure, calculated at a point $[v]\in \mathbb{P}(\mathcal{H})$, reads
\begin{gather}\label{eq:sympPURE}
\omega(\hat{\xi_1},\hat{\xi_2})=\frac{-i\bk{v}{[\xi_1,\xi_2]v}}{2\bk{v}{v}},
\end{gather}
whereas for $\mc O_\rho$ at a density matrix $\sigma$,
\begin{gather}\label{eq:sympMIXED}
\omega(\hat{\xi_1},\hat{\xi_2})=-\frac{i}{2}\tr (\sigma [\xi_1,\xi_2]),
\end{gather}
From the formulas \eqref{eq:sympPURE} and \eqref{eq:sympMIXED}  it is clear that the action of the full unitary group $SU(\mathcal{H})$  preserves the symplectic structures on $\mathbb{P}(\mathcal{H})$ and $\mathcal{O}_\rho$. The same concerns the action of compact subgroups $K$ of $SU(\mathcal{H})$. As explained in the Appendix, this fact can be used to define  
the \emph{momentum map} associated to the action of $K$, $\mu: M \rightarrow \mathfrak{k}$, where $M=\mathbb{P}(\mc H)$ or $M=\mathcal{O}_\rho$ (recall that $\mathfrak{k}=Lie(K)$). 
The momentum map  $\mu:\mathbb{P}(\mc H)\rightarrow \mathfrak{k}$ for
pure states of  $L$ distinguishable particles is given by
\begin{gather}\label{dist}
\mu([v])=\frac{i}{2}\{\rho_1([v])-\frac{1}{N_1}I_{N_1},\rho_2([v])-\frac{1}{N_2}I_{N_2},\ldots,\rho_L([v])-\frac{1}{N_L}I_{N_L}\},
\end{gather}
where $\rho_i([v])$ is the $i$-th  reduced one-particle density matrix
of a state $[v]\in \mathbb{P}(\mathcal{H})$\footnote{The one-particle density matrix on a mixed  $L$-particle state $\rho$ can be defined via the identity
\[
\mathrm{tr}(\rho_i A) = \mathrm{tr}\left( \rho I_{N_1} \otimes \ldots \otimes I_{N_{i-1}} \otimes  A \otimes I_{N_{i+1}} \otimes \ldots \otimes I_{N_{L}} \right)\ ,    
\]
which should hold {\it for all} operators $A$ acting on the space $\mathcal{H}_i$. A pure state $[v]\in \mathbb{P}(\mathcal{H})$ can be identified with rank one projector onto $\mathcal{H}$ i.e, $\rho([v])= \ket{v}\bra{v}/\bk{v}{v}$. The reduced density matrices in Eq.(\ref{dist}) should be computed exactly for this projector.} and  $I_{N_i}$ is the
identity operator on the $N_i$-dimensional Hilbert space $\mc H_i$. For
pure states of $L$ $d$-state indistinguishable particles (bosons or
fermions) the momentum map reads reads $\mu([v])=\frac{i}{2}(\rho_1([v])-\frac{1}{d}I)$,
$\rho_1([v])$. For the case of mixed states of distinguishable particles with the fixed spectrum, $\mc
O_\rho$, the momentum map is given by the formula similar to Eq.(\ref{dist}), but with $\rho_i$ being
now the reduced one-particle density matrices of a state $\sigma\in\mc
O_\rho$.

\section{Local unitary equivalence of quantum states \label{sec:skrot}}

In all cases considered above the momentum map relates a state to its
one-particle reduced density matrices. This observation paves a way of
relating local unitary equivalence to properties of the reduced
states or, in other words, to properties of the momentum map. Before discussing the usefulness of the momentum map in the context of the problem of LU equivalence let us briefly review the 'standard` approach to this problem based on the concept of invariant polynomials.

Since the symmetry group $K$ is a compact group, its orbits are themselves compact (and hence also closed). For this reason  in order to check if two  states belong to the same orbit, it suffices to check whether the values for all $K$-invariant polynomials\footnote{Invariant polynomials are $K$-invariant functions which are polynomials in the coordinates of the states.},  evaluated for these two states, are the same. Moreover, a celebrated theorem by Hilbert states that, for a compact group $K$ acting in a
unitary fashion on a finite-dimensional vector space, the ring of polynomial invariants is finitely generated \cite{M03}. Translating this to the physical problem in question, we conclude that there exists a finite number of independent invariant  polynomials that are able to distinguish whether two states are LU equivalent. The rings of polynomial invariants were computed for biparticle scenarios \cite{Grasl1998} as well as for the case of three qubits \cite{Sudbery2000}. Moreover, analogous ideas have been explored in the context of LU equivalence for mixed states \cite{Vrana2010,LUinv1} as well as LU equivalence of multipartite bosonic states  \cite{Migdal2014}. Despite the fact that polynomial invariants can be in principle measured (provided  having access to many copies of the state of interest \cite{InvMeasure}), their number increases drastically with the size of the system. What is more, values of invariant polynomials are in general hard to interpret and hence provide little physical insight into problem. We believe that the approach to the LU-equivalence based on the properties of momentum map, although somewhat limited, provides a new insight into this problem. Before we proceed let us note that in the literature there exist complementary approaches to LU-equivalence that do not explicitly use invariant polynomials. In particular, the work  \cite{Barbara1} derived a collection  necessary and sufficient (although difficult to check for greater number of particles)  conditions for LU-equivalence of multiquibit states.

 The equivariant momentum map, $\mu:M\rightarrow \mathfrak{k}$, maps
$K$-orbits in the space of states $M$ onto orbits of the adjoint action
of $K$ in $\mathfrak{k}$, the Lie algebra of $K$. Consequently, each
$\mathrm{Ad}_{K}$-invariant polynomial $p:\mathfrak{k}\rightarrow
\mathbb{R}$ on $\mathfrak{k}$, when composed with $\mu$, gives a
$K$-invariant polynomial $P=p\circ \mu :M\rightarrow \mathbb{R}$ on
$M$. The invariant polynomials for the adjoint action of $SU(N)$ are
generated by traces of powers not larger than $N$ of $X\in
\mathfrak{su}(N)$. Combining this with the fact that the momentum map
is given by reduced one-particle density matrices we conclude that
traces of their powers are $K$-invariant polynomials on $M$. If the
pre-image of every adjoint orbit from $\mu(M)\subset\mathfrak{k}$
consists of exactly one $K$-orbit in $M$, a $K$-orbit on $M$ (i.e., a
set of equally correlated states) can be identified by the traces of
powers of the reduced one-particle density matrices. Since the traces
of powers of a matrix determine its spectrum (and \textit{vice versa})
we can decide upon local unitary equivalence of states by examining
the spectra of their reduced one-particle density matrices, hence, in principle by measurements performed independently in each laboratory.

Alas, the situation described above is quite exceptional. Typically
many $K$-orbits are mapped onto one adjoint orbit. Hence, even if
one-particle density matrices of two states have the same spectra we
need additional $K$-invariant polynomials to decide their
$K$-equivalence \cite{SWK13}, \cite{MOS13}. The number of the
additional polynomials characterizes, in a certain sense, the amount of
additional information that can not be obtained from local
measurements, but needs to be inferred from non-local ones, i.e.,
involving effectively more than one subsystem. The whole problem can be
looked upon as a quantum version of the classical `marginal problem',
where we try to recover a probability distribution from its marginals
(here, recover a quantum state from its reduced density matrices)
\cite{K04}.

\begin{figure}[ht!]\label{fig:LU_transforms}
\centering \resizebox{0.45\linewidth}{!}{\includegraphics{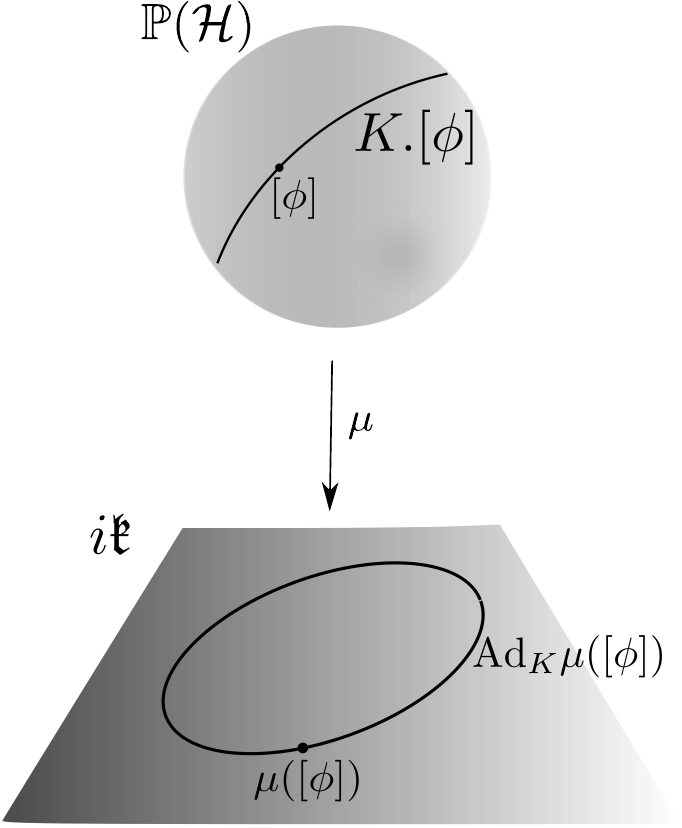}}
\caption{The ball represents the projective Hilbert space $\mathbb{P}(\mathcal{H})$ with the state $[\phi]$ and its K-orbit $K.[\phi] $. The orbit is transformed by the momentum map, $\mu:\mathbb{P}(\mathcal{H})\mapsto \mathfrak{k}$, to the orbit of the adjoint action $\mathrm{Ad}_K\mu([\phi])$ (see Eq. \ref{dist} for the definition of $\mu([\phi])$). Note that the canonical definition of $\mu$ is given by the coadjoint action $\mathrm{Ad}^\ast_g\zeta,\zeta\in\mathfrak{k}^\ast$ (more in Appendix), i.e. $\mu:\mathbb{P}(\mathcal{H})\mapsto \mathfrak{k}^\ast$. In this paper, however, we identify adjoint and coadjoint action. }
\end{figure}

Let us, however, concentrate for a moment on cases (exceptional, as noted in the above remarks), when we infer local unitary equivalence upon examining the spectra of the reduced states. 

\subsection{Spherical embeddings and determination of the local equivalence from the spectra of the reduced states}\label{sec:spherical}

As pointed above, the situation when the set of $K$-invariant polynomials on $M$ is given by the composition of $\Ad_K$-invariant polynomials on $\mathfrak{k}$ with the momentum map  $\mu:M\rightarrow \mathfrak{k}$ occurs only when the pre-image of every adjoint orbit from $\mu(M)\subset \mathfrak{k}$ is exactly one $K$-orbit in $M$. Let us thus consider the fibre of the momentum map $\mu$ over $\mu(x)\in \mathfrak{k}$,
\begin{equation}\label{eq:fiber}
\mathcal{F}_{x}:=\{z\in M:\mu(z)=\mu(x)\},
\end{equation}
i.e. the set of all points in $M$ that are mapped to the point $\mu(x)$ by the momentum map. The position of  $\mathcal{F}_x$ with respect to the orbit $K.x$ is of key importance for the above situation. The orbits of $K$ in $M$ are in 1-1 correspondence with the adjoint orbits in $\mu(M)$ if and only if each $\mc F_x$ are contained the orbit $K.x$, since in this case all points mapped to $\mu(x)$ are on the same orbit (that can be thus identified, when we know $\mu(x)$). In order to simplify the formulas in the discussion below, in what follows we will use the notation $K.x \equiv \mathcal{O}_x$.

It is easy to see that the tangent space $T_{y}\mathcal{F}_{x}$ at $y$ is $\omega$-orthogonal to the tangent space $T_y\mathcal{O}_x$ at the same point $y$, i.e., for each $a\in T_{y}\mathcal{F}_{x}$ and $b\in T_y\mathcal{O}_x$ we have $\omega(a,b)=0$. Indeed, let $\{\xi_k\}$, $k=1,\ldots,d=\dim\mathfrak{g}$,  be a basis in the Lie
algebra $\mathfrak{g}$. The corresponding vector fields $\hat{\xi}_k$ at $x$
(c.f.\ Eq.~(\ref{eq:fundvfield}) in the Appendix) span the tangent space $T_x\mathcal{O}_x$ to the
orbit through $x$ at $x$. On the other hand, the fiber $\mathcal{F}_x$ is a
common level set of the functions $\mu_{\xi_k}$,
\begin{equation}\label{levelset}
\mathcal{F}_x=\{z\in M:\mu_{\xi_k}(z)=c_k\},\quad c_k=\mu_{\xi_k}(x), \quad
k=1,\ldots,d,
\end{equation}
which means that the derivatives of $\mu_{\xi_k}$ vanish on the tangent to $\mathcal{F}_x$, $d\mu_{\xi_k}(a)=0$ for $a\in T_x\mathcal{F}_x$.

Define now the kernel of $d\mu$,
\begin{equation}\label{kermu}
\mathrm{Ker}_x(d\mu):=\{a\in T_x M:d\mu_{\xi_k}(x)(a)=0,k=1,\ldots,d\}.
\end{equation}
From the definition of the momentum map (see Appendix)  we have
\begin{equation}\label{dmuak}
d\mu_{\xi_k}(a)=\omega(\hat{\xi}_k,a).
\end{equation}
Hence $a\in\mathrm{Ker}_x(d\mu)$ if and only if $a$ is $\omega$-orthogonal to
all $\hat{\xi}_k$. Since  $\hat{\xi}_k$ span $T_x\mathcal{O}_x$ we have
\begin{equation}\label{f1}
\mathrm{Ker}_x(d\mu)=(T_x\mathcal{O}_x)^{\perp\omega},
\end{equation}
where by $X^{\perp\omega}$ we denote the space of $\omega$-orthogonal vectors to $X$. Combining this with the observation above that $d\mu_{\xi_k}$ vanish on  $T_x\mathcal{F}$, we
find that $T_x\mathcal{F}_x\subset \mathrm{Ker}_x(d\mu)$ and, consequently,
$T_x\mathcal{F}_x\subset(T_x\mathcal{O}_x)^{\perp\omega}$. The above reasoning clearly does not depend on the choice of a
particular point in $\mathcal{F}_x$, i.e., as announced,
\begin{equation}\label{fact1}
T_y\mathcal{F}_x\subset(T_y\mathcal{O}_x)^{\perp\omega}, \quad y\in\mathcal{F}_x.
\end{equation}

A submanifold $P$ of a symplectic manifold $M$ is called coisotropic if for
arbitrary $y\in P$ we have $(T_yP)^{\perp\omega}\subset T_yP$. We conclude
thus that if $\mathcal{O}_x$ is coisotropic then
$\mathcal{F}_x\subset\mathcal{O}_x$. Indeed from (\ref{fact1}) and the
coisotropy of $\mathcal{O}_x$ at each $y\in\mathcal{F}_x$ we have
$T_y\mathcal{F}_x\subset T_y\mathcal{O}_x$. Hence, in this case examining
whether some $y$ belongs to $\mathcal{O}_x$ (and, consequently whether $y$
and $x$ are LU-equivalent) reduces to checking whether their corresponding one-particle reduced states have the same spectra \cite{SK11}.

In order to characterize all systems for which LU-equivalence can be decided using the spectra of one-particle density matrices we need to identify those whose (at least) generic $K$-orbit is coisotropic. As it turns out, such systems need to satisfy some group theoretic conditions \cite{HKS13}.   To reach the final solution we have to study not only the action of $K$ but also of its complexification $G= K^\mathbb{C}$ on $M$. 
%Note that since $K$ is a maximal compact subgroup of $G$, the group $G$ is {\it reductive}. An important example of the reductive group is $G=\mathrm{SL}_{N}(\mathbb{C})$, that is complexification of its compact subgroup  $K=\mathrm{SU}(N)$. By this example groups $G$ considered in [6] (and introduced in section \ref{subsec:moment}) are reductive. 
Note that the group $G$ is much bigger than $K$ and therefore the number of $G$-orbits is  smaller than the number of $K$-orbits in $M$. If $G$ has an open dense orbit on $M$ then we call $M$ \textit{almost homogenous manifold} \cite{huckleberry90}.  The almost homogenity of $M$ with respect to the $G$-action is a necessary condition for deciding $K$-equivalence on $M$ using the momentum map \cite{HKS13}. It is, however, not a sufficient condition, what can be seen from the example of the three-qubit system, where there are exactly $6$ orbits of  $G=SL(2,\mathbb{C})^{\times 3}$, but the states  $x_{1}=\sqrt{\frac{2}{3}}\ket{000}+\frac{1}{\sqrt{3}}\ket{111}$ and $x_{2}=\frac{1}{\sqrt{3}}\left(\ket{000}+\ket{010}+\ket{001}\right)$, where $\{\ket{0},\,\ket{1}\}\subset\mathbb{C}^2$ is an orthonormal basis in $\mathbb {C}^2$, satisfy $\mu(x_{1})=\mu(x_{2})$ but are not $K$-equivalent as they belong to different $G$-orbits \cite{dur00}.

An important role in the formulation of the sufficient condition is played by the Borel subgroup of the group $G$. By definition a Borel subgroup $B$ is a maximal connected solvable subgroup of the group $G$. For example, for $G=\mathrm{SL}_{N}(\mathbb{C})$, the group of upper-triangular matrices (of unit determinant) is an example of a Borel subgroup. This subgroup of $\mathrm{SL}_{N}(\mathbb{C})$ is a stabilizer of the standard full flag in $\mathbb{C}^N$, (i.e., a collection of subspaces with the dimension growing by one, such that the given one includes all subspaces with smaller dimensions),
\[
0\subset\mathrm{Span}\{e_1\}\subset\mathrm{Span}\{e_1,e_2\}\subset\ldots\subset\mathrm{Span}\{e_1,\ldots,e_{N-1}\}\subset\mathrm{Span}\{e_1,\ldots, e_N\}=\mathcal{H}\,
\]
where $e_1,\ldots,e_N$ is a basis in $\mathcal{H}$. Generally, any two Borel subgroups are conjugated by an element of $G$. Therefore, in the considered example,  $B$ is a Borel subgroup of $G$ if and only if it stabilizes some standard full flag. 

The crucial notion for the $K$-equivalence problem is the notion of a {\it spherical space}. If $G$ is a group and $H$ its subgroup we call $\Omega=G/H$, i.e., the space obtained from $G$ by identifying points connected by elements of $H$, a \textit{G-homogenous space}. In this space the group $G$ acts in a natural way, simply by the group product followed by identification of elements connected by elements of $H$, hence we can consider actions of $G$ or its subgroups on $\Omega$, in particular of a Borel subgroup. This leads to the following definition. A $G$-homogenous space  $\Omega=G/H$ is a \textit{spherical homogenous space} if and only if  some (and therefore every)  Borel subgroup $B\subset G$ has an open dense orbit in  $\Omega$. 

Now observe that an orbit of $G$ in the initially considered manifold $M$ through a point $x$ can be identified with some $G$-homogenous space. Indeed we can take as $H$ the subgroup that stabilizes $x$, than all other points on the orbit are obtained by actions of the elements of  $G$ that effectively move $x$. 

All these considerations allow the following definition. If $G$ has an open dense orbit $\Omega=G/H$ in $M$ and $\Omega$ is a spherical homogenous space, then $M$ is called a \textit{spherical embedding} of $\Omega=G/H$. Such $M$ is also called \textit{almost homogenous spherical space} (with respect to the action of $G$). Its relevance stems from the following Brion's theorem \cite{brion87},
\begin{theorem}\label{brion result}(Brion) Let $K$ be a connected
	compact Lie group acting on connected compact K\"{a}hler manifold
	$(M,\omega)$ by a Hamiltonian action and let $G=K^{\mathbb{C}}$. The
	following are equivalent
	\begin{enumerate}
		\item $M$ is a spherical embedding of the open $G$-orbit.
		\item For every $x\in M$ the fiber $\mathcal{F}_x$ is contained
		in $K.x$.
	\end{enumerate}
\end{theorem}
In other words, the second point above assures that the momentum map separates all $K$-orbits, i.e., from $\mu(x) = \mu(y)$ it follows $y\in K.x$.
In our case the group $K$ is the local unitary group, i.e., the group of the product of the local unitaries. The conclusion is that, if the relevant $\mathbb{P}(\mathcal{H})$ (the space of states of the whole system) is a spherical embedding of an open orbit of the corresponding group of the product of the local special linear groups (i.e.\ the complexification of the local unitary group), then LU-equivalence can be decided upon spectra of the reduced one-particle states. 

As it should be clear from the reasoning presented above we need to find an open dense orbit of $G$ in $\mathbb{P}(\mathcal{H})$  and show that $\mathbb{P}(\mathcal{H})$ is a spherical embedding of it (in short, that the action of $G$ on $\mathbb{P}(\mathcal{H})$ is \textit{spherical}). In \cite{HKS13} we use Brion's theorem and show that open dense orbits of the Borel subgroup exist for systems of two fermions, two bosons and two distinguishable particles in arbitrary dimensions.   Therefore, for such system the LU-equivalence can be decided using reduced one-particle density matrices. Interestingly, exactly these systems have been studied previously in the context of quantum information \cite{CORNOND2002,Nielsen1999}.

\subsection{Exceptional states and tensor rank}

From the previous section we conclude that typically we are confronted with undecidability of local unitary equivalence upon examining the spectra of reduced one-particle density matrices. The problem can be be looked upon from another point of view, using tools borrowed from algebraic geometry, by exhibiting causes for such a situation.  In particular, one can show that an obstacle is the existence of so-called exceptional states. To describe them we define the rank \cite{land12}  of a state in $M$. The rank of a state is defined with respect to Perelomov coherent states  $\mathbb{X}$ of a particular $K$-action on $\mathbb{P}\left( \mathcal{H} \right)$ \cite{perel}.  For  our discussion it will be important that the variety $\mathbb{X}$ consists of coherent states that are ''closest to classical''. This means that $\mathbb{X}$ can be understood as the orbit of $K$ through the state corresponding to the highest-weight vector in $\mathcal{H}$. In the cases considered by us the variety $\mathbb{X}$ can be also viewed as the image of the following three maps, corresponding to distinguishable particles, bosons and fermions, respectively,   

\begin{enumerate}
\item The Segre map, $\mathrm{Seg}_L:\PP(\mc
    H_1)\times...\times\PP(\mc H_L) \lw \PP \left(\mc H_1\otimes\dots\mc H_L\right)$, $
   \mathrm{Seg}_L([v_1], ...,[v_L]) \lo [v_1\otimes ...\otimes v_L]
   $. The image is the set of separable states.

\item The Veronese map, $\mathrm{Ver}_L : \PP(\mc H_1)  \lw \PP\left(S^LH_1\right)$,
    $\mathrm{Ver}_L[v] \lo [v^{\otimes L}]$. The image is the set of spin coherent states for the angular momentum operator or, more generally, permamental states of bosons.

\item The Pl\"ucker map, $\mathrm{Pl}_L : Gr(L,\mc H_1) \lw \PP\left(\Lambda^LH_1\right)$, where $Gr(L,\mc H_1)$ is the Grassmanian, i.e., the space
    of all $L$-dimensional subspaces of $\mc H_1$,
    $\mathrm{Pl}_L\big(\mathrm{span}\{u_1,\ldots u_L\}\big) \lo
    [u_1\wedge...\wedge u_L]$. The image is the set of Slater determinantal states for the fixed number of fermions.
\end{enumerate}

The rank of a state with respect to $\mathbb{X}$ is defined as
\begin{equation}
{\rm rk}[\psi] = {\rm rk}_\XX[\psi] = \min \{r\in\NN : \psi={x}_1+\dots
+{x}_r, \  [x_j]\in\XX\}\,,
\end{equation}
where $\XX$ is the image of the corresponding map (1.-3. above). In words, it means that the rank is the minimal number of states needed to obtain the state in question from states in $\mathbb{X}$.  Let us remark that in the case of two particles the notions of the rank introduced above correspond to the standard matrix rank (for distinguishable particles), the rank of symmetric forms (for bosons) and finally the rank of skew-symmetric forms (for fermions).

By definition, the rank of a state is invariant under the action of group $G$ (and hence also under the action of $K$). For the case of two distinguishable particles the concept of  tensor rank  has been used to define a more fine-grained classiffication of quantum entanglement of mixed states \cite{SR1,SR2} (see also \cite{MC1, MS1} for the usage of this concept in the context of SLOCC transformations between pure multipartite states)

The sets of states of rank $r$ will be denoted by $\XX_r = \{
[\psi]\in\PP(\mc H) : {\rm rk}[\psi] = r \}$. They are not closed\footnote{in the Zariski topology} and it turns out that there are states in $\PP(\mc
H)$ of a certain rank $r$ that can be approximated with an arbitrary
precision by states of a lower rank. 

An example of an exceptional state is a 3-qubit $W$-state of rank 3,
$\ket{W}=\frac{1}{\sqrt{3}}(\ket{011}+\ket{101}+\ket{110})$. It can be
obtained as the limit of a sequence of rank-2 states obtained from the
GHZ state, $\ket{GHZ}=\frac{1}{\sqrt{2}}(\ket{000}+\ket{111})$. Indeed
in $\PP(\mc H)$ we have,
\begin{equation}
A(a)^{\otimes 3}[GHZ] \stackrel{a\rw 0}{\lw}[W], \quad
A(a) =\frac{1}{\sqrt{2}} \begin{pmatrix} a & a \\ -a^{-1} & a^{-1} \end{pmatrix}
\end{equation}

It was proved \cite{ST13} that their
existence is an obstacle preventing from inferring the local unitary
equivalence of states from the image of the momentum map, i.e., from
the spectra of the reduced one-particle density matrices.

The main theorem of \cite{ST13} reads:
\begin{theorem}\label{Theo Spher<->no except}
	Suppose that we have one of the following three configurations of a state space ${\mc H}$, a complex reductive Lie group $G$ acting irreducibly on ${\mc H}$, and a variety of coherent states $\XX\subset\PP(\mc H)$, which is the unique closed $G$-orbit in the projective space $\PP(\mc H)$.
	
	{\rm (i)} ${\mc H_D}=\mc H_1\otimes ...\otimes \mc H_L$, $G_D=GL(\mc H_1)\times ...\times GL(\mc H_L)$, $\XX=\mathrm{Segre}(\PP(\mc H_1)\times...\times\PP(\mc H_L))$. 
	
	{\rm (ii)} ${\mc H_B}=S^L(\mc H_1)$, $G=GL(\mc H_1)$, $\XX=\rm{Ver}_L(\PP(\mc H_1))$.
	
	{\rm (iii)} ${\mc H_F}=\bigwedge^L\mc H_1$, $G=GL(\mc H_1)$, $\XX=\rm{Pl}(Gr(L,\mc H_1))$.
	
	\noindent Then the action of $G$ on ${\PP(\mc H_{B,F}})$ (resp. $G_D$ on $\PP(\mc H_{D}$)) is spherical if and only if there are no exceptional states in $\PP(\mc H_{B,F})$ (resp. $\PP(\mc H_{D})$) with respect to $\XX$. In other words, sphericity of the representation is equivalent to the property that states of a given rank cannot be approximated by states of lower rank.
\end{theorem}
Combining this theorem with the results described in the previous section we infer that the existence of exceptional states is an obstacle for deciding LU-equivalence using the momentum map i.e., upon reduced spectra.

\subsection{General case - how many independent parameters are needed to decide LU-invariance?}

As already mentioned above, information about the spectra of the
reduced one-particle density matrices does not allow to decide whether
two states of the whole system belong to the same orbit of the local
unitary transformation group, i.e.\ whether they have the same correlation
properties. The condition $\mu(K.[\psi])=\mu(K.[\phi])$, %where
%$K.[\phi]$ denotes the orbit in $\mathbb{P}(\mathcal{H})$ of $K$
%through $[\psi]$, etc.,
 is still a necessary one for the
$K$-equivalence of states $[\phi]$ and $[\psi]$. The image of the
momentum map, $\mu(M)$ consists of adjoint orbits in $\mk{k}$. Each
adjoint orbit intersects the Cartan subalgebra $\mk{t}$ (maximal
commutative subalgebra of $\mk{k}$  at a finite number of points. This
statement expresses the fact that each matrix from $\mk{k}$ can be
diagonalized by the adjoint action (conjugation) of an element of the
group $K$ (just like a (anti-)Hermitian matrix can be diagonalized by
some unitary transformation). The Cartan subalgebra, in this case, is
represented by diagonal matrices, but the eigenvalues (diagonal
elements at the point of intersection) are ordered in some specific
way. By permutations we can order them differently, reaching in this
way some other point of intersection. Thus different intersection
points are connected by the action of some subgroup of the permutation
group - the Weyl group. To make the situation unambiguous we impose a
concrete order, for example nonincreasing, which means that we choose
the intersection point that belongs to a subset of diagonal matrices
(i.e., elements of $\mathfrak{t}$ with nonincreasing diagonal entries),
called the positive Weyl chamber and denoted by $\mathfrak{t}_+$.  Now
we can define $\Psi:M\rightarrow\mathfrak{t}_+$ to be the map
satisfying $\Psi([\phi])=\mu(K.[\phi])\cap\mathfrak{t}_{+}$. It assigns
to a state $[\phi]$ the ordered spectra of the (shifted) one-particle
reduced density matrices. By taking the intersection of the whole image
of the momentum map, $\mu(M)$, with the positive Weyl chamber,
$\mathfrak{t}_+$, one obtains the set
$\Psi(M)=\mu(M)\cap\mathfrak{t}_+$ parametrizng the orbits by elements
of $\mk{t}_+$ \cite{guillemin84}. The convexity theorem of the momentum
map \cite{Atiyah82,GS84,K84} states that $\Psi(M)$ is a convex
polytope, referred to as the Kirwan polytope. 

The necessary condition for states $[\phi_1]$ and $[\phi_2]$ to be $K$-equivalent, can be therefore formulated as $\Psi([\phi_1])=\Psi([\phi_2])$. To decide whether they are really equivalent we need to inspect additional invariants. eg., additional invariant polynomials. It is shown in \cite{MOS13} that for $L$-qubit states satisfying the necessary condition, the number of additional invariant polynomials strongly depends of the spectra of the one-qubit reduced density matrices, i.e., on the point in the polytope $\Psi(M)$. For $\alpha\in \Psi(M)$, the number of additional polynomials is given by by the dimension of the reduced space $M_\alpha=\Psi^{-1}(\alpha)/K$. In \cite{MOS13} $\rm{dim}_\RR\Psi^{-1}(\alpha)/K$ was analyzed for arbitrary $\alpha\in \Psi(M)$.  

For multi-qubit systems the inequalities describing the polytope $\Psi(M)$ are known \cite{HSS03}. Denote by $\{p_i,1-p_i\}$ an increasingly ordered spectrum of the $i$-th reduced density matrix and by $\lambda_i$ the shifted spectrum,
$\lambda_i=\frac{1}{2}-p_i$. Then, $\Psi(M)$ is given by
$0\leq\lambda_{l}\leq\frac{1}{2}$ and
$\left(\frac{1}{2}-\lambda_{l}\right)\leq\sum_{j\neq
l}\left(\frac{1}{2}-\lambda_{j}\right)$. Methods used in
\cite{MOS13} to compute the dimensions of spaces $M_\alpha$, are
different for points $\alpha$ belonging to the interior of $\Psi(M)$
and for points from the boundary of the polytope. For more than two
qubits, the polytope is of full dimension, hence a generic $K$-orbit in
the space of states $M$ has the dimension of $K$ \cite{SWK13}. Using
the regularity of $\mu$ \cite{HH96, MW99} we get that for points
$\alpha$ from the interior of the polytope the dimension of the reduced
space reads:
\begin{gather}
\mathrm{dim}M_{\alpha}=\mathrm{dim}(\Psi^{-1}(\alpha)/K)=\left(\mathrm{dim}\mathbb{P}(\mathcal{H})-\mathrm{dim}\Psi(\mathcal{H})\right)-\mathrm{dim}K=\nonumber \\
=\left(\left(2^{L+1}-2\right)-L\right)-3L=2^{L+1}-4L-2.\label{eq:fiber-interior-1}
\end{gather}
Points belonging to the boundary of $\Psi(M)$ can be grouped into three
classes: (i) $k$ of $\lambda_{l}$ are equal to $\frac{1}{2}$, (ii) at
least one of inequalities
$\left(\frac{1}{2}-\lambda_{l}\right)\leq\sum_{j\neq
l}\left(\frac{1}{2}-\lambda_{j}\right)$ is an equality, (iii) $k$ of
$\lambda_{l}$ are equal to $0$. In case $(i)$, inequalities that yield
$\Psi(M)$ reduce to an analogical set of inequalities for the
$(L-k)$-qubit polytope. Therefore,
$\mathrm{dim}M_{\alpha}=\left(\left(2^{L-k+1}-2\right)-(L-k)\right)-3(L-k)=2^{L-k+1}-4(L-k)-2$.
States that are mapped to points that fall into case (ii) belong to the
$K^\mathbb{C}$-orbit through the $L$-qubit $W$-state,
$[W]=\ket{01\ldots1}+\ket{101\ldots1}+\ldots+\ket{1\ldots10}$
\cite{MOS13}. As it is shown in \cite{SWK13}, the closure of such an
orbit is an almost homogeneous spherical space. Therefore, the fibers of
the momentum map are contained in $K$-orbits (see Section~\ref{sec:spherical}),
i.e.\,$\mathrm{dim}M_\alpha =0$. Case (iii), where $k$ of $\lambda_{l}$
are equal to $0$, is the most difficult one, as it requires the use of
some more advanced tools from the Geometric Invariant Theory (GIT)
\cite{M03}. Here a key role is played by {\it stable
states} \cite{M03,M77}, i.e.\,states for which $\mu([\phi])=0$ and
$\mathrm{dim}K.\ket{\phi}=\mathrm{dim} K$ \cite{Kirwan-thesis}. For a
symplectic action of a compact group $\tilde K$ the existence of stable
states implies that
$\tilde{\mu}^{-1}(0)/\tilde{K}=\mathrm{dim}\mathbb{P}(\mathcal{H})-2\mathrm{dim}\tilde{K}$,
where $\tilde \mu$ is the momentum map for the $\tilde K$-action. The
strategy taken in \cite{MOS13} for case (iii) is the following. The group $K$ can be divided into $K=K_{1}\times K_{2}$,
where $K_{1}=SU(2)^{\times k}$, $K_{2}=SU(2)^{\times(L-k)}$ and $K_1$
acts on the first $k$ qubits.  The action of $K_1$ yields
the momentum map, which assigns to a state its first $k$ one-qubit
reduced density matrices. Therefore, $\mu_{1}^{-1}(0)$ consists of
states whose first $k$ reduced density matrices are maximally mixed,
while the remaining $(L-k)$ reduced matrices are arbitrary. Further one  constructs a state that is GIT stable with respect
to the action of $K_1^\mathbb{C}$ on $\mathbb{P}(\mathcal{H})$. Hence,
$\mathrm{dim}\mu_{1}^{-1}(0)/K_{1}=\mathrm{dim}\mathbb{P}(\mathcal{H})-2\mathrm{dim}K_{1}=2^{L+1}-6k-2$.
Furthermore, the quotient $\mu_{1}^{-1}(0)/K_{1}$ is a symplectic
variety itself. Because the actions of $K_1$ and $K_2$ commute, we can
consider the action of $K_2$ on $\mu_{1}^{-1}(0)/K_{1}$. The momentum
map for $K_{2}$ acting on $\mu_{1}^{-1}(0)/K_{1}$ gives the remaining
$L-k$ one-qubit reduced density matrices. The polytope of $\Psi_{2}$ is
of full dimension, i.e.\,of dimension $L-k$. By the formula for the
dimension of the reduced space for points from the interior of the
polytope, we get:
\begin{gather*}
\left(\left(\mathrm{dim}\mu_{1}^{-1}(0)/K_{1}\right)-\left(L-k\right)\right)-\mathrm{dim}K_{2}=\left(\left(2^{L+1}-6k-2\right)-\left(L-k\right)\right)-3(L-k)\\
=2^{L+1}-4L-2k-2,
\end{gather*}
which is the desired result for the case (iii).
\begin{figure}[H]
\centering
\includegraphics[width=17cm]{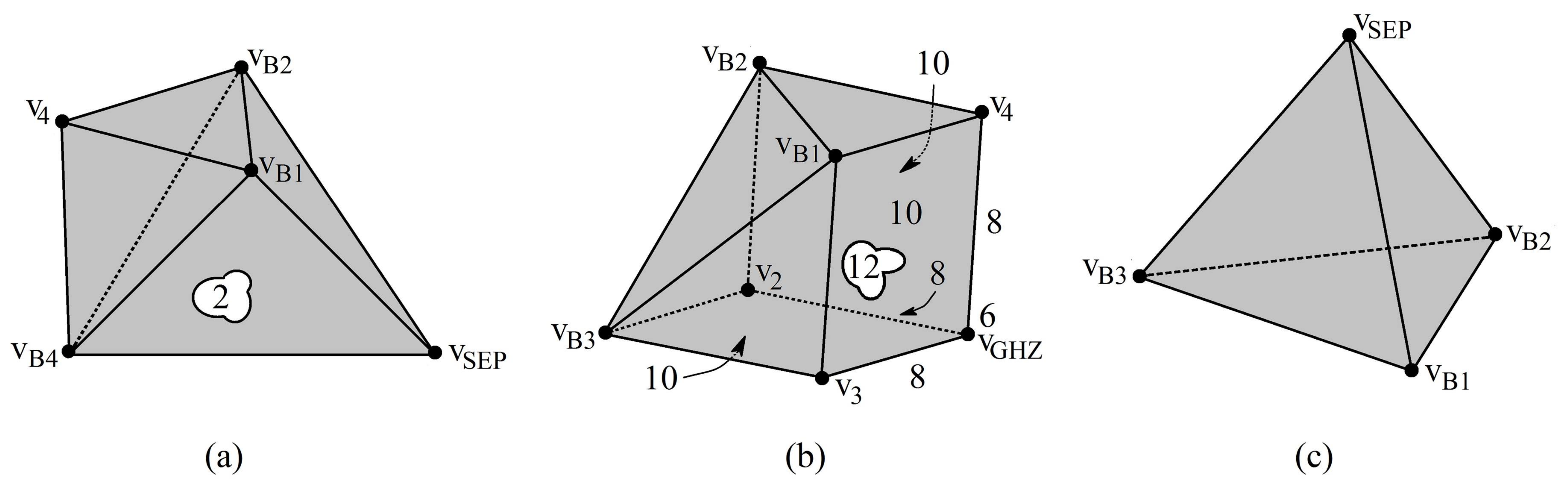}
\caption{\label{fig:Three-kinds-of}The three parts of the boundary of $\Psi(\mathbb{P}(\mathcal{H}))$ for four qubits. The numbers denote $\mathrm{dim}M_{\alpha}$.
If the number is missing, then $\mathrm{dim}M_{\alpha}=0$.} % Taken from \cite{MOS13}.}
\end{figure}

\section{SLOCC-equivalence of quantum states}\label{sec:kcrow}
A classification of states with respect to SLOCC \cite{Vidal} can be, as already
mentioned, treated as a complementary one to the LU-equivalence
characterization of quantum correlations. For reasons
briefly explained below such a classification is not an easy task and
its intricacies are still not fully understood. To a large extend the
arising problems were fairly exhaustively explained in \cite{SOK14}.

\subsection{Invariant polynomials approach}
For the considered multipartite systems reversible SLOCC operations
correspond to elements of the complexification $G=K^\mathbb{C}$ of the
local unitary group $K$, and two states are SLOCC
equivalent if and only if they belong to the same $G$-orbit. Recall
that the problem of $K$-equivalence is solvable by means of
$K$-invariant polynomials. As the group $G$ is reductive, (which means
that it is a complexification of its maximal compact group - in this
case the compact group $K$), the Hilbert-Nagata theorem \cite{M03}
ensures that the ring of $G$-invariant polynomials is finitely
generated, just like in the compact case of the local unitary group
$K$. Nevertheless, the problem of $G$-equivalence turns out to be
significantly different from the problem of $K$-equivalence. The essence of this difference is the fact that the
group $G$ is not compact and thus $G$-orbits do not have to be closed.
For two vectors $\phi_1$ and $\phi_2$ on two non-intersecting orbits,
the closures of the orbits can intersect. Since $G$-invariant
polynomials are continuous functions, they are not able to distinguish
between orbits $ G.\phi_1$ and $G.\phi_2$ in such a case. It is only
possible to distinguish between orbits whose closures have non-empty
intersection, in particular between closed $G$-orbits (a complete solution of the SLOCC-equivalence problem in this special case have been obtained in \cite{Wallach1}). In purely
topological language this `pathology' can be linked to the fact that
the orbit space $M/G$, i.e., the quotient space identifying states on
the same orbit, is not longer a Hausdorff space - not every pair of
points is separated by open sets. The $G$-equivalence of states is thus
intimately linked to the structure of the orbit space resulting from
the action of a non-compact reductive group on a vector space $\mc H$
(equivalently on the projective space $\mathbb{P}(\mc H)$).

%\begin{figure}[ht!]\label{fig:slocc_pic}
%\centering\resizebox{0.4\linewidth}{!}{\includegraphics{slocc.pdf}}
%\caption{An example of SLOCC-equivalent space. The picture presents the Hilbert space with the states $|\phi_1\rangle,|\phi_2\rangle$ and their G-orbits. The dashed lines denote the open orbits $G.\phi_1$ and $G.\phi_2$ that have zero intersection. The solid lines represent the closures of the orbits and their intersection is nonempty. }
%\end{figure}

\subsection{Geometric Invariant Theory approach}\label{sec:GITa}
Two orbits of $ G.\phi$ and $G.\psi$ in $\mathcal{H}$ are called
$c$-equivalent if and only if there exists a sequence of orbits
$G.\phi=G.v_1,\, G.v_{2},\ldots,\, G.v_{n}=G.\psi$ such that the
closures of each two consecutive ones intersect,
$\overline{G.v_{k}}\cap\overline{G.v_{k+1}}\neq\emptyset$. The relation
of $c$-equivalence divides $G$-orbits into equivalence classes
($c$-classes). It turns out that every $c$-class contains exactly one
closed $G$-orbit contained in the closure of every $G$-orbit belonging
to the considered $c$-class. The equivalence classes are thus
parametrized by closed $G$-orbits and $G$-invariant polynomials
distinguish between $G$-orbits belonging to different $c$-classes
\cite{M03}. Among all $c$-classes we distinguish those corresponding to
the zero vector - they form the so-called null cone \cite{Ness84}. This
class must be removed if we want to consider the quotient space at the
projective level. After removing from the projective space
$\mathbb{P}(\mc H)$ points corresponding to vectors from the null cone
we are left with so called {\it semistable} points, $\mathbb{P} (\mc
H)_ {ss}$. Two points $x_{1},x_{2}\in\mathbb{P}(\mathcal{H})_{ss}$ are
$c$-equivalent  if there are vectors of $v_{1},v_2\in\mathcal{H}$, such
that $x_{1}=[v_{1}]$ and $x_{2}=[v_{2}]$ and on the level of the
Hilbert space $\overline{G.v_{1}}\cap\overline{G.v_{2}}\neq\emptyset$.
The quotient space obtained from the semistable points by
$c$-equivalence relation is denoted by
$\mathbb{P}(\mathcal{H})_{ss}\sslash G$ and is a projective algebraic
variety. It is known in the literature under the name \emph{GIT
quotient} \cite{M77}. Points of the
GIT quotient correspond to $c$-classes of semistable points and are in
one-to-one correspondence with closed $G$-orbits. It turns out that
every closed $G$-orbit in $\mathbb{P}(\mc H)_{ss}$ contains exactly one
$K$-orbit from $\mu^{-1}(0)$ \cite{KN82}. Therefore we get the
following equivalence
$\mu^{-1}(0)/K\cong\mathbb{P}(\mathcal{H})_{ss}\sslash G$, giving thus
a description the GIT quotient in terms of the momentum map (see Figure~4)

% remove
\begin{figure}[h]\label{fig:GIT}
\centering 
\includegraphics[scale=0.37]{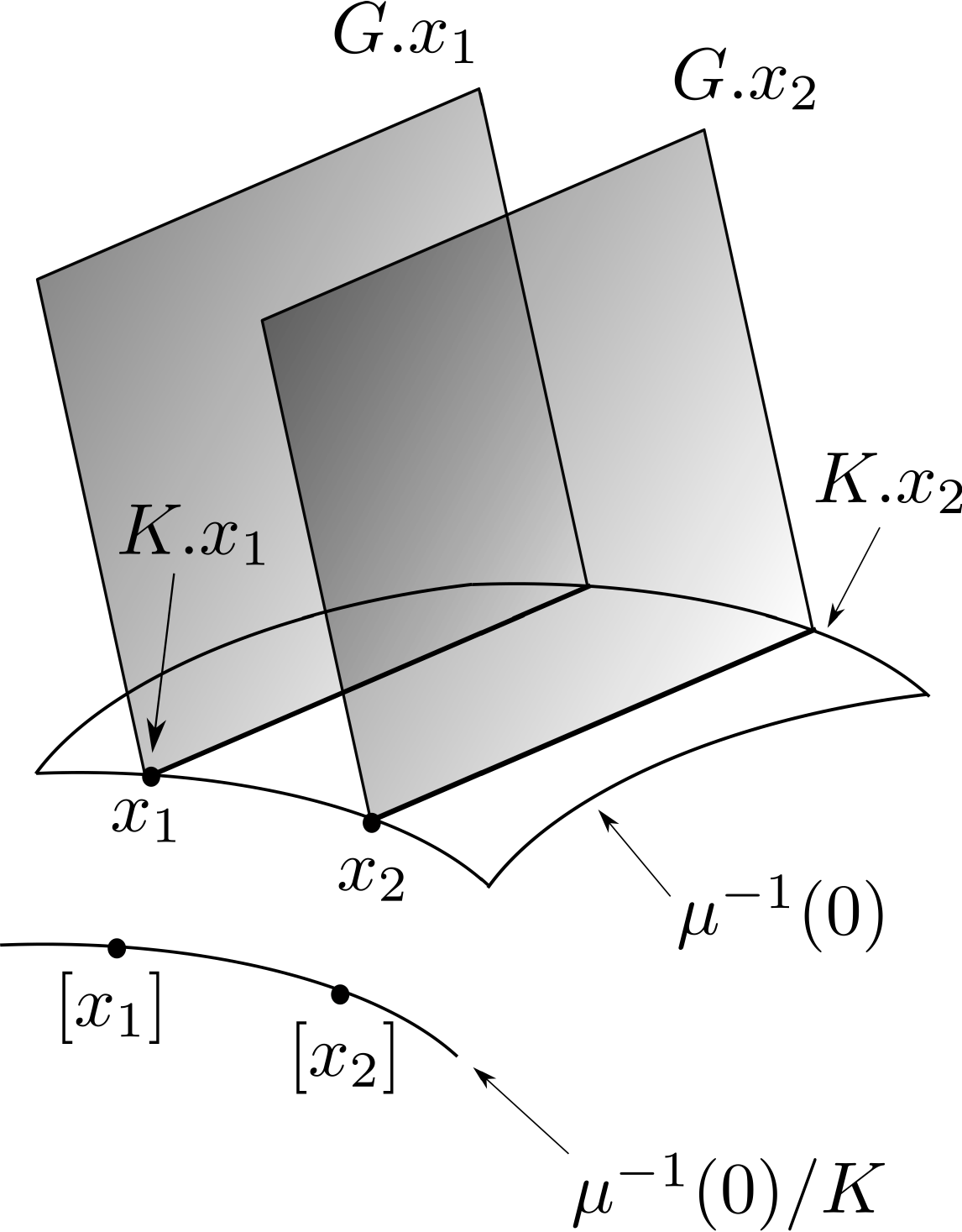}

\caption{The idea of the GIT quotient  construction, i.e.
$\mu^{-1}(0)/K\cong\mathbb{P}(\mathcal{H})_{ss}\sslash G$.} % Taken from
%\cite{SOK14}.}
\end{figure}

The set of closed $G$-orbits in $\mathbb{P}(\mathcal{H})_{ss}$ is given by the action of $G$ on
$\mu^{-1}(0)$, i.e. $G.\mu^{-1}(0)$. Among the semistable points we
distinguish the so-called stable points
$\mathbb{P}(\mathcal{H})_{s}=\{x\in\mathbb{P}(\mathcal{H})_{ss}:\,\dim
G.x=\dim
G\,\,\mbox{and}\,\,\mbox{\ensuremath{G.x\cap\mu^{-1}(0)\neq\emptyset}}\}$
\cite{M77}. The existence of a single stable point makes
$\mathbb{P}(\mathcal{H})_{s}$ an open dense subset of
$\mathbb{P}(\mathcal{H})_{ss}$, i.e. almost every semistable point is
stable \cite {Milne05}. For the stable point $x\in
\mathbb{P}(\mathcal{H})_{s}$ the $c$-equivalence class consists of
exactly one closed $G$-orbit. For  semistable but not stable points
this class always consists of an infinite number of $G$-orbits.

Vectors belonging to the null cone, i.e. $c$-class whose closed
$G$-orbit is the zero vector, may represent important states from the
point of view of quantum correlations. For example, the W-state
$\ket{W}=1/\sqrt{3}(\ket{110}+\ket{101}+\ket{011})$ and separable
states belong to the null cone but their quantum properties are
significantly different. Therefore we need a finer procedure dividing
$G$-orbits, one that includes the GIT construction and also provides
mathematically and physically well-defined stratification of the null
cone. A key role is played here by the function
$||\mu||^2:\mathbb{P}(\mc H)\rightarrow\mathbb{R}$, i.e., the norm of
the momentum map. It has a clear mathematical and physical
interpretation. According to the definition given by Klyachko
\cite{klyachko07},  the total variance of state
$[v]\in\mathbb{P}(\mathcal{H})$ with respect to the symmetry group
$K\subset SU(\mathcal{H})$ is given by
\begin{gather}
\mathrm{Var}([v])=\frac{1}{\bk vv}\left(\sum_{i=1}^{\dim K}\bra v\xi_{i}{}^{2}\ket v
-\frac{1}{\bk vv}\sum_{i=1}^{\dim K}\bra v\xi_{i}\ket v^{2}\right)=c-4\cdot\left\Vert \mu\right\Vert
^{2}([v]),
\label{eq:variance}
\end{gather}
where ${\xi_i}$ form an orthonormal basis of algebra $\mathfrak{k}$ and
$c$ is a $[v]$-independent constant. The function $\left\Vert
\mu\right\Vert ^{2}([v])$ can be also expressed as the expectation
value of the Casimir operator \cite{BR80},
$\mathcal{C}_{2}=\sum_{i=1}^{\dim K}\xi_{i}^{2}$ for the irreducible
representation of $K$ on the symmetrized tensor product
$\mathrm{Sym}^2\mathcal{H}$ \cite{oszmaniec12}. In this case we have
$\mathcal{C}_{2}^{\vee}=\sum_{i=1}^{\dim K}(\xi_{i}\otimes
I+I\otimes\xi_{i})^{2}$ and $\frac{1}{\bk vv^{2}}\bra{v\otimes
v}C_{2}^{\vee}\ket{v\otimes v} =2c+8\left\Vert \mu\right\Vert
^{2}(\left[v]\right)$. Finally, $\|\mu\|^2$ is directly related to the
linear entropy, which is a linear function of  the total variance.

A point $[v]\in \mathbb{P}([v])$ is a critical point of $\|\mu\|^2$ if
it is a solution of  $\mu([v]).v=\lambda v$ \cite{SOK14}. Critical
points of $\|\mu\|^2$ can be therefore divided into two categories. The
first includes all $K$-orbits belonging to $\mu^{-1}(0)$. These are
called minimal critical points and for them $\|\mu\|^2$ reaches a
global minimum. The minimal critical points correspond to states with
maximum total variance and maximum linear entropy. The other critical
points are given by some $K$-orbits in the null cone. For these points
$\mu([v])\neq 0$ and $\mu([v])v=\lambda v$. Therefore, in  the null
cone we distinguish $G$-orbits passing through the critical $K$-orbits.

The relationship between critical points of $\|\mu\|^2$,
$c$-equivalence and GIT construction becomes clear if we consider the
gradient flow of $-\|\mu\|^2$ \cite{Ness84}. The gradient of
$-\|\mu\|^2$ is well defined as the projective space $\mathbb{P}(\mc
H)$ is a K\"ahler manifold, and therefore is equipped with a well
defined metric determined by its Riemannian structure (see Appendix). The gradient
flow is tangent to $G$-orbits and carries points towards critical
$K$-orbits. Two points $x_1,x_2\in\mathbb{P}(\mc H)_{ss}$ are
equivalent from the point of view of the gradient flow if they are
taken by it to the same critical $K$-orbit. This definition is
consistent with the $c$-equivalence definition. However, it is at the
same time more general because it allows an extension of the concept of
equivalence to the null cone. The situation in the null cone is more
complex as the critical $K$-orbits do not need to be in one fiber of
$\Psi$ (recall that $\Psi:\mathbb{P}(\mc H)\rightarrow\mathfrak{k}$ is
given by $\Psi(\ket{\phi})=\mu(K.\ket{\phi})\cap\mathfrak{t}_{+}$).
Nevertheless, the polytope $\Psi(\mathbb{P}(\mc H))$ has a finite
number of points $\{\alpha_i\}$ for which $\Psi^{-1}(\alpha_i)$
contains critical $K$-orbits. Let $C_\alpha$  denote the set of
critical $K$ orbits mapped by $\Psi$ on $\alpha \in \mathfrak{t}_+$ and
$N_\alpha$ be the set of all the points that are taken by gradient flow
of $-\|\mu\|^2$ to $C_\alpha$. The quotient space $N_\alpha\sslash G$,
obtained from $N_\alpha$ by dividing $N_\alpha$ by the equivalence
relation induced from the gradient flow and the space $C_\alpha/K$ are
isomorphic algebraic varieties (see Figure~5). 

\begin{figure}[H]
	~~~~~~~~~~~~~~~~~~~~~~~~~~~~~\includegraphics[scale=0.45]{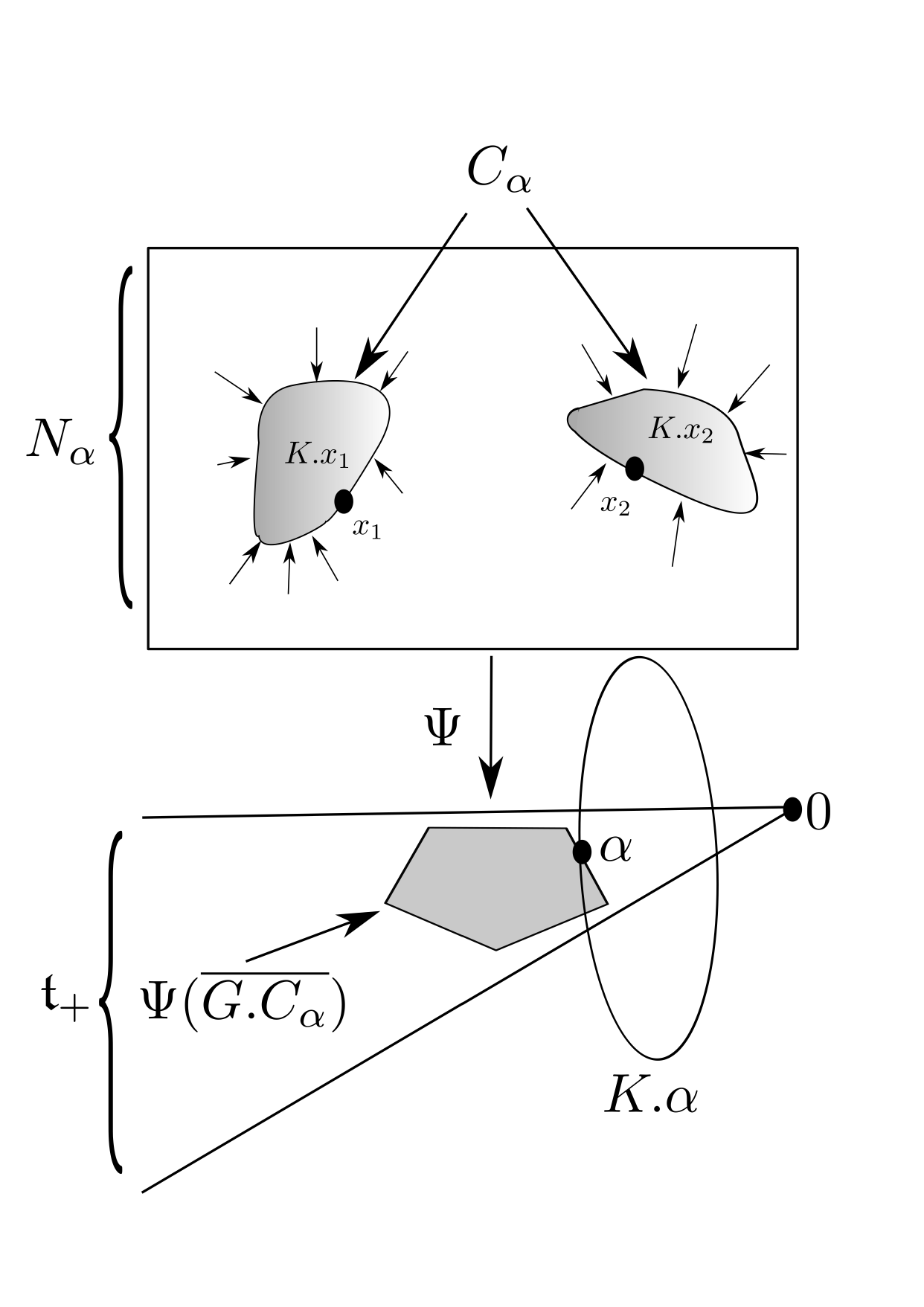}
	
	\caption{The sets $N_\alpha$ and $C_\alpha$, with two exemplary
		critical $K$-orbits, $K.x_1$ and $K.x_2$. The arrows represent the
		gradient flow of $-||\mu||^{2}$.} % Taken from \cite{SOK14}.}
\end{figure}

The above described
construction is thus analogous to the GIT one. For $\alpha=0$  we get
that $N_0 =\mathbb{P}(\mc H)_{ss}$ and $C_0=\mu^{-1}(0)$. Using the so
defined equivalence relation we can think of a quotient space
$\mathbb{P}(\mathcal{H})$ by $G$, i.e, the space identifying points on
the same $G$-orbit, as of the space consisting of a finite number of
projective algebraic varieties:
\begin{gather}\label{decom}
\mathbb{P}(\mathcal{H})/G\cong\bigcup_{\alpha}C_{\alpha}/K.
\end{gather}
In the above formula we abused slightly the notation writing $\mathbb{P}(\mathcal{H})/G$, since, as explained above this is not a `good quotient' from the point of view of the $G$-action.

The map $\Psi$ has another important property, namely not only
$\Psi(\mathbb{P}(\mc H))$ is a convex polytope but also the image of
every $G$-orbit closure $\Psi(\overline{G.x})$ has this property
\cite{brion87}. A finite number of varieties $C_{\alpha}/K$ is the
result of the fact that  $N_\alpha$ can be equivalently defined as
those $x\in\mathbb{P}(\mathcal{H})$ for which polytopes
$\Psi(\overline{G.x})$ share the nearest point to the origin. But the
momentum map convexity theorem for $G$-orbits ensures that the number
of such polytopes is finite \cite{GuliSjamaar06}, so the number of
manifolds $C_\alpha$ is also finite. Summarizing, we obtained the correspondence given in Table~1.
\begin{table}[H]
	\centering
	\caption{A dictionary }\label{table}
	\begin{tabular}{|p {0.25\linewidth}|p {0.6\linewidth}|}
		\hline
		$G$-orbit & SLOCC class of states \tabularnewline
		\hline
		the momentum map $\mu$ & the map which assigns to a state $[v]$ the collection of its reduced one-particle density matrices\tabularnewline
		\hline
		$||\mu||^2([v])$ & the total variance of state $\mathrm{Var}([v])$, linear entropy \tabularnewline
		\hline
		closure equivalence class of orbits & family of asymptotically equivalent SLOCC classes  \tabularnewline
		\hline
		stable point & SLOCC family consists of exactly one SLOCC class \tabularnewline
		\hline
		semistable but not stable point & SLOCC family consists of many SLOCC classes \tabularnewline
		\hline
		$\Psi(\overline{G.[v]})$ & SLOCC momentum polytope, collection of all possible spectra of reduced one-particle density matrices for $[u]\in\overline{G.[v]} $ \tabularnewline
		\hline
		strata $N_\alpha$ & group of families of SLOCC classes - all states for which SLOCC momentum polytopes have the same closest point to the origin\tabularnewline
		\hline
		$C_\alpha$ &  set of critical points of $\mathrm{Var}([v])$ with the same spectra of reduced one-particle density matrices \tabularnewline
		\hline
	\end{tabular}
\end{table}

In this way we achieved
decomposition into a \underline{finite} number of SLOCC classes
determined by a single, easily accessible function
$||\mu||^2$, which can be expressed in terms of the total variance of
the state. In comparison with the approach using invariant polynomials
the above presented one shows considerable advantages. It
differentiates between states that clearly differ with respect to their
correlation properties e.g., the W-state and separable states, which the
invariant polynomial method puts to the same class. Moreover, invariant
polynomials usually do not have clear physical meaning, in general they
are not experimentally accessible, in contrast to the total variance,
which can be measured. And finally, the number of different classes is
finite, what enables an effective classification.

The space $\mathbb{P}(\mc H)$ can be also divided into a finite number
of generalized SLOCC classes using the polytopes $\Psi(\overline{G.x})$,
which in  \cite{CDKW12} are called entanglement polytopes. This is done
by saying that two states are equivalent when their entanglement
polytopes are the same. Decomposition (\ref{decom}) is identical with
that division up to the existence of polytopes that have a common
closest point to the origin.

The key ingredient needed to obtain the decomposition (\ref{decom}) is the
knowledge of the critical $K$-orbits of $\|\mu\|^2$. In \cite{SOK14} they were found for two distinguishable and indistinguishable particles,
three qubits and any number of two-state bosons. For four qubits it was shown that most classes found in \cite{V03} are $c$-equivalent with the
class corresponding to $\mu^{-1}(0)$.

\begin{table}[H]
	\centering
	\begin{tabular}{|c|c|c|}
		\hline
		Critical $\alpha\in\Psi(\mathbb{P}(\mathcal{H}))$ & State & E($\phi$)\\
		\hline
		\hline
		$\left(\begin{array}{cc}
		-\frac{1}{2} & 0\\
		0 & \frac{1}{2}
		\end{array}\right),\,\left(\begin{array}{cc}
		-\frac{1}{2} & 0\\
		0 & \frac{1}{2}
		\end{array}\right),\,\left(\begin{array}{cc}
		-\frac{1}{2} & 0\\
		0 & \frac{1}{2}
		\end{array}\right)$  & Sep & 0\\
		\hline
		$\left(\begin{array}{cc}
		-\frac{1}{2} & 0\\
		0 & \frac{1}{2}
		\end{array}\right),\,\left(\begin{array}{cc}
		-\frac{1}{2} & 0\\
		0 & \frac{1}{2}
		\end{array}\right),\,\left(\begin{array}{cc}
		0 & 0\\
		0 & 0
		\end{array}\right),\,\left(\begin{array}{cc}
		0 & 0\\
		0 & 0
		\end{array}\right)$ &  TriSep & $\frac{1}{4}$\\
		\hline
		$\left(\begin{array}{cc}
		-\frac{1}{6} & 0\\
		0 & \frac{1}{6}
		\end{array}\right),\,\left(\begin{array}{cc}
		-\frac{1}{6} & 0\\
		0 & \frac{1}{6}
		\end{array}\right),\,\left(\begin{array}{cc}
		-\frac{1}{6} & 0\\
		0 & \frac{1}{6}
		\end{array}\right),\,\left(\begin{array}{cc}
		-\frac{1}{2} & 0\\
		0 & \frac{1}{2}
		\end{array}\right)$ & $\ket{\textrm{W}^{(3)}}\otimes\ket{1}$ & $\frac{1}{3}$\\
		\hline
		$\left(\begin{array}{cc}
		-\frac{1}{2} & 0\\
		0 & \frac{1}{2}
		\end{array}\right),\,\left(\begin{array}{cc}
		0 & 0\\
		0 & 0
		\end{array}\right),\,\left(\begin{array}{cc}
		0 & 0\\
		0 & 0
		\end{array}\right),\,\left(\begin{array}{cc}
		0 & 0\\
		0 & 0
		\end{array}\right)$ &  BiSep & $\frac{3}{8}$\\
		\hline
		$\left(\begin{array}{cc}
		-\frac{1}{4} & 0\\
		0 & \frac{1}{4}
		\end{array}\right),\,\left(\begin{array}{cc}
		-\frac{1}{4} & 0\\
		0 & \frac{1}{4}
		\end{array}\right),\,\left(\begin{array}{cc}
		-\frac{1}{4} & 0\\
		0 & \frac{1}{4}
		\end{array}\right),\,\left(\begin{array}{cc}
		-\frac{1}{4} & 0\\
		0 & \frac{1}{4}
		\end{array}\right)$ &  W & $\frac{3}{8}$\\
		\hline
		$\left(\begin{array}{cc}
		-\frac{1}{10} & 0\\
		0 & \frac{1}{10}
		\end{array}\right),\,\left(\begin{array}{cc}
		-\frac{1}{10} & 0\\
		0 & \frac{1}{10}
		\end{array}\right),\,\left(\begin{array}{cc}
		-\frac{1}{5} & 0\\
		0 & \frac{1}{5}
		\end{array}\right),\,\left(\begin{array}{cc}
		-\frac{1}{5} & 0\\
		0 & \frac{1}{5}
		\end{array}\right)$ &  $\Phi_3$ & $\frac{9}{20}$\\
		\hline
		$\left(\begin{array}{cc}
		-\frac{1}{6} & 0\\
		0 & \frac{1}{6}
		\end{array}\right),\,\left(\begin{array}{cc}
		-\frac{1}{6} & 0\\
		0 & \frac{1}{6}
		\end{array}\right),\,\left(\begin{array}{cc}
		-\frac{1}{6} & 0\\
		0 & \frac{1}{6}
		\end{array}\right),\,\left(\begin{array}{cc}
		0 & 0\\
		0 & 0
		\end{array}\right)$ & $\Phi_2$ & $\frac{11}{24}$\\
		\hline
		$\left(\begin{array}{cc}
		-\frac{1}{14} & 0\\
		0 & \frac{1}{14}
		\end{array}\right),\,\left(\begin{array}{cc}
		-\frac{1}{14} & 0\\
		0 & \frac{1}{14}
		\end{array}\right),\,\left(\begin{array}{cc}
		-\frac{1}{14} & 0\\
		0 & \frac{1}{14}
		\end{array}\right),\,\left(\begin{array}{cc}
		-\frac{1}{7} & 0\\
		0 & \frac{1}{7}
		\end{array}\right)$ &  $\Phi_1$ & $\frac{27}{56}$\\
		\hline
		$\left(\begin{array}{cc}
		0 & 0\\
		0 & 0
		\end{array}\right),\,\left(\begin{array}{cc}
		0 & 0\\
		0 & 0
		\end{array}\right),\,\left(\begin{array}{cc}
		0 & 0\\
		0 & 0
		\end{array}\right),\,\left(\begin{array}{cc}
		0 & 0\\
		0 & 0
		\end{array}\right)$ &  GHZ & $\frac{1}{2}$\\
		\hline
	\end{tabular}
	\caption{One-qubit reduced density matrices for critical states of four
		qubits. The listed states are:
		$\ket{\textrm{TriSep}}=\frac{1}{\sqrt{2}}\ket{11}\otimes
		(\ket{00}+\ket{11})$,
		$\ket{\textrm{BiSep}}=\frac{1}{\sqrt{2}}\ket{1}\otimes
		(\ket{000}+\ket{111})$,
		$\ket{\textrm{W}}=\frac{1}{2}(\ket{1110}+\ket{1101}+\ket{1011}+\ket{0111})$,
		$\ket{\Phi_3}=\sqrt{\frac{3}{10}}(\ket{1101}+\ket{1110})+\sqrt{\frac{2}{5}}\ket{0011}$,
		$\ket{\Phi_2}=\frac{1}{2\sqrt{3}}(\ket{1011}+\ket{1110})-\frac{1}{2}(\ket{0101}+\ket{0011})+\frac{1}{\sqrt{3}}\ket{0110}$,
		$\ket{\Phi_1}=\sqrt{\frac{3}{14}}(\ket{0011}+\ket{0101}+\ket{1001})+\sqrt{\frac{5}{14}}\ket{1110}$,
		$\ket{\textrm{GHZ}}=\frac{1}{\sqrt{2}}(\ket{0000}+\ket{1111})$.}
	\label{tab:4qubit_old}
\end{table}

\subsection{Critical points of $\|\mu\|^2$ for many qubits}

The critical points of $||\mu||^2$, or of the linear entropy, play a
key role in understanding the generalized SLOCC classes. Finding the
critical states by direct application of the definition is a
computationally difficult task, as it requires solving the equation
$\mu([v]).v=\lambda v$, i.e.\ finding eigenvectors and eigenvalues of a
matrix depending nonlinearly on a vector it acts on, thus rendering 
seemingly straightforward eigenvectors-eigenvalues problem effectively
nonlinear. In \cite{MS15} a slightly more tractable method on an
interplay between momentum maps for abelian and non-abelian Lie groups
was proposed.

For a compact group $K$, we denote by $T$ its maximal torus, which is a
maximal connected abelian subgroup. For example, when
$K=SU(N)$, a maximal torus consists of unitary diagonal matrices with
the determinant one. A momentum map $\mu_T:M\rightarrow \mathfrak{t}$
for the action of $T$ on $M$ is given by the composition of
$\mu:M\rightarrow \mathfrak{k}$ with the projection on the Cartan
subalgebra $\mathfrak{t}=Lie(T)$. Therefore, we have
$\mu([v])=\mu_T([v])+\alpha$, where $\alpha\in \mathfrak{t}^\bot$. By
the convexity theorem,  $\mu_T(M)$ is a convex polytope. For abelian
groups, the convexity theorem specifies the vertices of the polytope
\cite{Atiyah82}. The vertices are among the elements of the set of weights
$\mathbb{A}=\mu_T(M^T)$, where $M^T$ are the fixed points for the
action of $T$ on $M$\footnote{A point $x\in M$ is fixed by the action
of $T$ iff $\forall t\in T\,t.x=x$}. The critical points of
$\|\mu_T\|^2$, must satisfy a similar condition as the critical points
of $\|\mu\|^2$, i.e. $\mu_T([v]).v=\lambda v$. Therefore, for $\beta
\in \mu_T(M)$ a point $[v]$ is a critical point if and only if a) $\beta .v=\lambda v$, i.e. $[v]$ is a fixed point for $T_\beta=\{e^{t\beta}:t\in
\mathbb{R}\}$ and b) $\mu_T([v])=\beta$. Following \cite{Kirwan-thesis} we
define $Z_\beta$ to be the set of those $[v]\in M^{T_\beta}$ that satisfy $\langle\mu_T([v]),\beta\rangle=\langle\beta,\beta\rangle$. One shows that 
$Z_\beta$ is a symplectic variety \cite{Kirwan-thesis}. Moreover, it is easy to see that points from $Z_\beta$ are sent by $\mu_T$ to the hyperplane
that is perpendicular to $\beta$ and contains $\beta$ \cite{Kirwan-thesis}. The set $Z_\beta$ is a $T$-invariant symplectic variety, hence, by the
convexity theorem, we have that $\mu_T(Z_\beta)$ is a convex polytope,
which is spanned by a subset of weights from $\mathbb{A}$. The
definition of $Z_\beta$ implies that $\beta$ is the closest to zero
point of this polytope. In other words, $[v]\in M$ is a critical point
of $\|\mu_T\|^2$ if and only if it is mapped to a minimal convex combination of weights, $\beta$, and $[v]\in Z_\beta$ \cite{Kirwan-thesis}.
\begin{figure}[h]
\centering
\includegraphics[width=.7\textwidth]{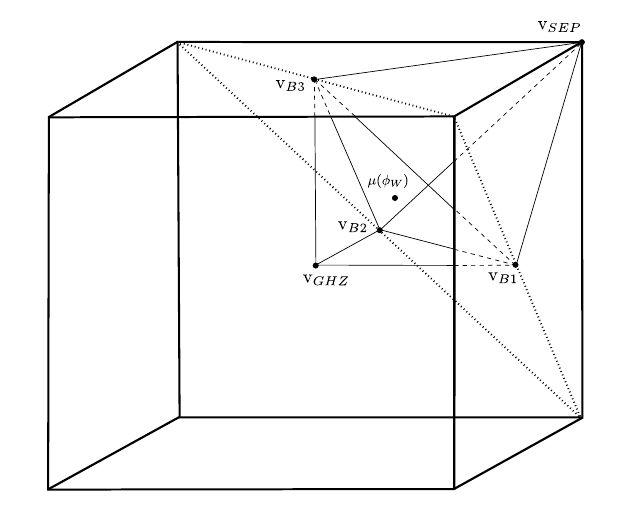}
\caption{Minimal weight combinations for three qubits. The point $\mathrm{v}_{GHZ}$ is the image of the state $\ket{GHZ}=\frac{1}{\sqrt{2}}\left(\ket{000}+\ket{111}\right)$, the points $\mathrm{v}_{Bi}$ correspond to the biseparable states and $\ket{\phi_W}=\frac{1}{\sqrt{3}}\left(\ket{110}+\ket{101}+\ket{011}\right)$.} % Taken from \cite{MS15}. }
\label{fig:3qubit}
\end{figure}

ThefFunction $\|\mu\|^2$ is $K$-invariant, therefore we can restrict our
consideration to critical points satisfying $\mu([v])\in
\mathfrak{t}_+$. For such states we have $\mu([v])=\mu_T([v])$ and
$[v]$ is a critical point of $\|\mu\|^2$ iff it is a critical point of
$\|\mu_T\|^2$. Let us denote by $\mathcal{B}$ the set of all minimal
combinations of weights from $\mathbb{A}$ that belong to
$\mathfrak{t}_+$. Then, a state $[v]$ is a critical one if and only if
$\mu([v])\in \mathcal{B}$ and $[v]\in Z_\beta$. Critical sets are
therefore of the form $C_\beta=K.(Z_\beta\cap\mu^{-1}(\beta))$, where
$\beta \in \mathcal{B}$. In \cite{MS15} the above
reasoning was applied to compute the critical points of the linear entropy
for pure states of $L$ qubits. The set $\mathbb{A}$ is the image under
$\mu$ of the basis states $B=\ket{i_1,\ldots,i_L}$, where
$i_k\in\{0,1\}$, hence $\#\mathbb{A}=2^L$. We discuss the algorithm of
finding the minimal combinations of weights and list the results up to
$L=5$ (the construction of the set of minimal combinations of weights
is shown on Figure 6). We also show that for $\beta\in \mathcal{B}$,
the set $Z_\beta=\mathbb{P}(\mathcal{S})$, where $S$ is spanned by the
basis states whose weights span $\beta$. Moreover, we show when sets
$C_\beta$ are nonempty and for each $\beta\in\mathcal{B}$ we describe a
construction of a state that is mapped to $\beta$. We conclude that the
number of critical values of the linear entropy grows
super-exponentially with $L$.

\section{Geometric and topological characterization of $CQ$ and $CC$ sates} \label{sec:cccq}

In previous sections we discussed applications of the momentum map in
two significant problems of the theory of quantum correlations, LU and SLOCC equivalence of pure states. In \cite{OSS14} it is shown how methods of symplectic and algebraic geometry can be applied to some concrete problems involving mixed states, namely to exhibit geometric and topological aspects of quantum correlations for separable (non-entangled) states.  The existence of quantum correlations for multipartite
separable mixed states can be regarded as one of the most interesting
quantum information discoveries of the last decade. In 2001 Ollivier
and \.{Z}urek  \cite{OZ01} and independently Henderson and Vedral
\cite{HV01} introduced the notion of quantum discord as a measure of
the quantumness of correlations. Quantum discord is always non-negative
\cite {D11}. The states with vanishing quantum discord are called {\it
pointer states}. They form the boundary between classical and quantum
correlations \cite {D11}. Bipartite pointer states can be identified
with the so-called classical-quantum, $CQ$ states \cite {D11}. An
important subclass of $CQ$ states are classical-classical, $CC$ states
\cite{KHH12}.

For $\mathcal{H}=\mathcal{H}_{A}\otimes\mathcal{H}_{B}$, where
$\mathcal{H}_{A}=\mathbb{C}^{N_{1}}$ and
$\mathcal{H}_{B}=\mathbb{C}^{N_{2}}$, a state is $CC$ if it can be
written as 
\begin{equation}
\rho=\sum_{i,j}p_{ij}\kb ii\otimes\kb jj\ ,
\end{equation}
 where real numbers $\lbrace p_{ij} \rbrace$ from a probability distribution,  $\{\ket
i\}_{i=1}^{N_{1}}$ is an orthonormal basis in $\mathcal{H}_{A}$ and
$\{\ket j\}_{j=1}^{N_{2}}$  is an orthonormal basis in $\mathcal{H}_{B}$. A state $\rho$ is  a
$CQ$ state if it can be written as 
\begin{equation}
\rho=\sum_{i}p_{i}\kb
ii\otimes\rho_{i}\ ,
\end{equation}
where numbers $\lbrace p_i \rbrace$ form a probability distribution and $\{\rho_{i}\}_{i=1}^{N_{2}}$ are the density
matrices on $\mathcal{H}_{B}$. Both $CC$ and $CQ$ states are of measure zero in $\mathcal{D}(\mc H)$ \cite{A10}. Importantly, for pure states the separable states are exactly the zero-discord states. It was shown in \cite{SHK11} that pure separable states are geometrically distinguished in the state space and belong to the unique symplectic $K$-orbit in $\mathbb{P}(\mc H)$. For mixed
states, already for two particles it is easy to see that there are
infinitely many symplectic $K$-orbits and there are separable states
through which  $K$-orbits are not symplectic. Thus a simple extension
of the results of \cite{SHK11}, even for two-particle mixed states is
not possible. In \cite{OSS14} four facts concerning geometric and
topological characterizations of $CC$ and $CQ$ states are shown. They extend results of \cite{SHK11} to mixed states: (1) the set of $CQ$ states is
the closure of all symplectic orbits of  $K=SU(N_1)\times I_{N_2}$, (2) the set of $CC$ states is the closure of all symplectic
orbits of $K=SU(N_1)\times SU(N_2)$, (3) the set of $CQ$ states
is exactly the set of $K=SU(N_1)\times I_{N_2}$ orbits whose
Euler-Poincar\'e characteristics $\chi$ do not vanish, (4) the set of
$CC$ states is exactly the set of $K=SU(N_1)\times SU(N_2)$ orbits
whose Euler-Poincar\'e characteristics $\chi$  do not vanish.

The space of all density matrices is not a symplectic space (the
symplectic form is degenerate). Nevertheless, as already mentioned in the introduction (see also Appendix), the set of density
matrices with the fixed spectrum  $\mc O_\rho$, which is the adjoint
orbit of  $SU(\mc H)$ through $\rho$ is symplectic. Therefore, the
action of the above given groups $K$ on $\mc O_\rho$ leads to existence
of the momentum map $\mu:\mc O_\rho\rightarrow \mathfrak{k}$
\cite{guillemin84}. In order to check if a given orbit $K.\sigma$ (the
action of $K$ on $\sigma\in\mc O_\rho$ is the adjoint action) is or is
not symplectic it is enough to consider the restriction of the momentum
map $\mu$ to $K.\sigma$. Then $K.\sigma$ is symplectic if this
restriction is bijective. The computational conditions for $\mu$ to be
bijective are  given in the Kostant-Sternberg theorem \cite{KS82}. Let us note that since $K.\rho$ is mapped
by $\mu$ onto an adjoint orbit in $\mathfrak{k}$, non-symplecticity of
$K.\rho$ (the degeneration of  symplectic form on $K.\rho$) can be
measured by $\mathrm{D}(K.\rho)=\dim K.\rho -\dim Ad_K\mu(\rho)$. For two qubits
the $CC$ states, in a fixed basis, form a 3-dimensional simplex and
therefore it is possible to see how the closure of the union of symplectic
$K=SU(N_1)\times SU(N_2)$ orbits forms the set of $CC$  states (Figure.
7, 8 and 9). In \cite{OSS14} we also discuss existence of  K\"ahler structure
and show that it is present on all considered symplectic $K$-orbits.

To find Euler-Poincar\'e characteristics $\chi$ we use the
Hopf-Samelson theorem \cite{HS41}. This theorem says that for action of
a compact group $K$ on a manifold $M$ the Euler-Poincar\'e
characteristics $\chi$ of the orbit $K.x$ passing through $x\in M$ is can be computed as follows
\begin{enumerate}
\item the  If the maximal torus $T$ of  $K$ is contained in $K_x$\footnote{Here, by $K_x$ we denote the stabiliser subgroup of $x$ in $K$.} then
    $\chi(K/K_x)=\frac{|W_K|}{|W_{K_x}|}$, where $W_K$ and
    $W_{K_x}$ are Weyl groups of $K$ and $K_x$ respectively.
\item Otherwise, $\chi(K/K_x)=0$.
\end{enumerate}
In \cite{OSS14} it is shown that orbits of the discussed groups through
$CC$ and $CQ$ states are the only orbits with stabilizer subgroups
containing maximal torus. The orders of the Weyl groups are calculated
and a formula for $\chi$ is given.

\begin{figure}[H]
\begin{centering}
\includegraphics[width=8cm]{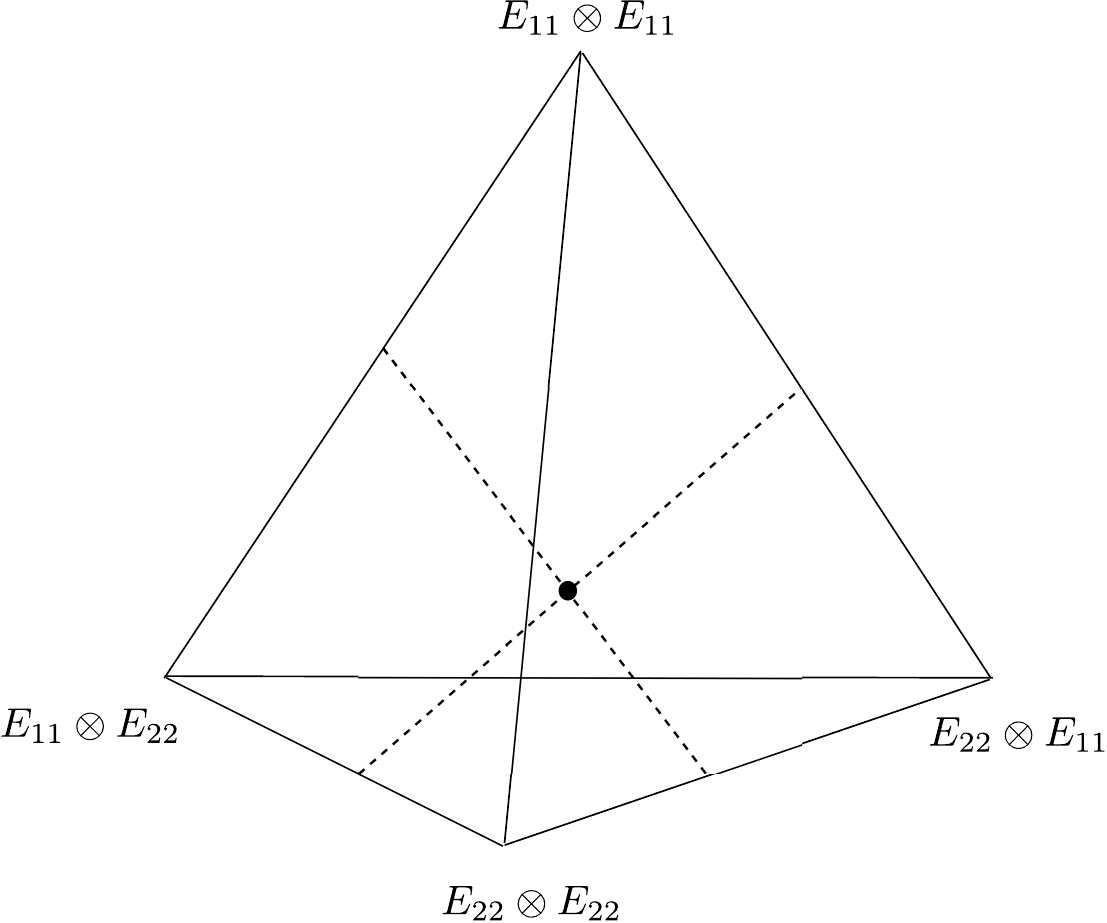}
\par\end{centering}

\caption{Dimensions of orbits through $CC$ states of two qubits. The large dot:
$\dim\, K.\rho=0$, the dotted lines: $\dim\, K.\rho=2$, elsewhere: $\dim\, K.\rho=4.$.} % Taken from \cite{OSS14}.}
\end{figure}

\begin{figure}[H]
\begin{centering}
\includegraphics[width=8cm]{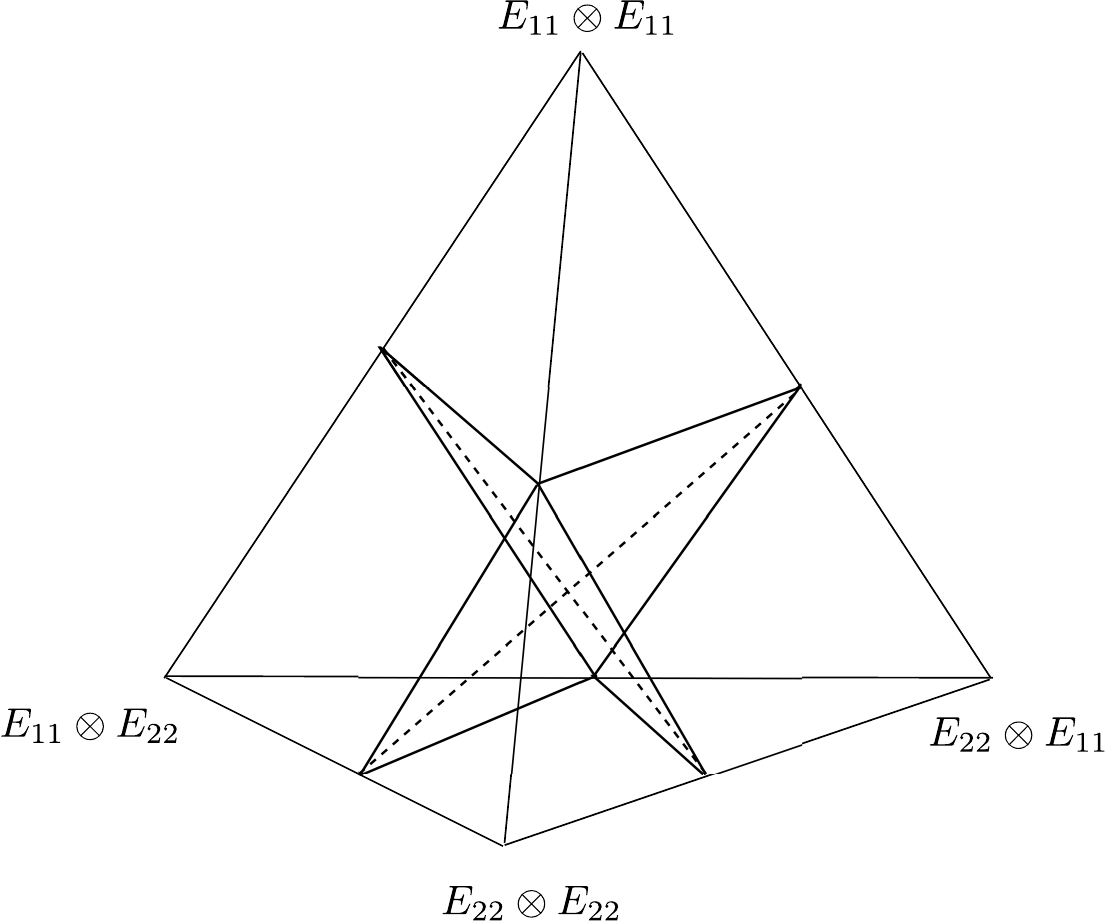}
\par\end{centering}

\caption{Ranks of $\left.\omega\right|_{K.\rho}$ for orbits through $CC$
states of two qbits. The thick dashed line: $\mathrm{rk}\left.\omega\right|_{K.\rho}=0$, the surfaces between thick solid lines: $\mathrm{rk}\left.\omega\right|_{K.\rho}=2$, elsewhere: $\mathrm{rk}\left.\omega\right|_{K.\rho}=4$.} %Taken from \cite{OSS14}.}
\end{figure}

\begin{figure}[H]
\begin{centering}
\includegraphics[width=0.5\linewidth]{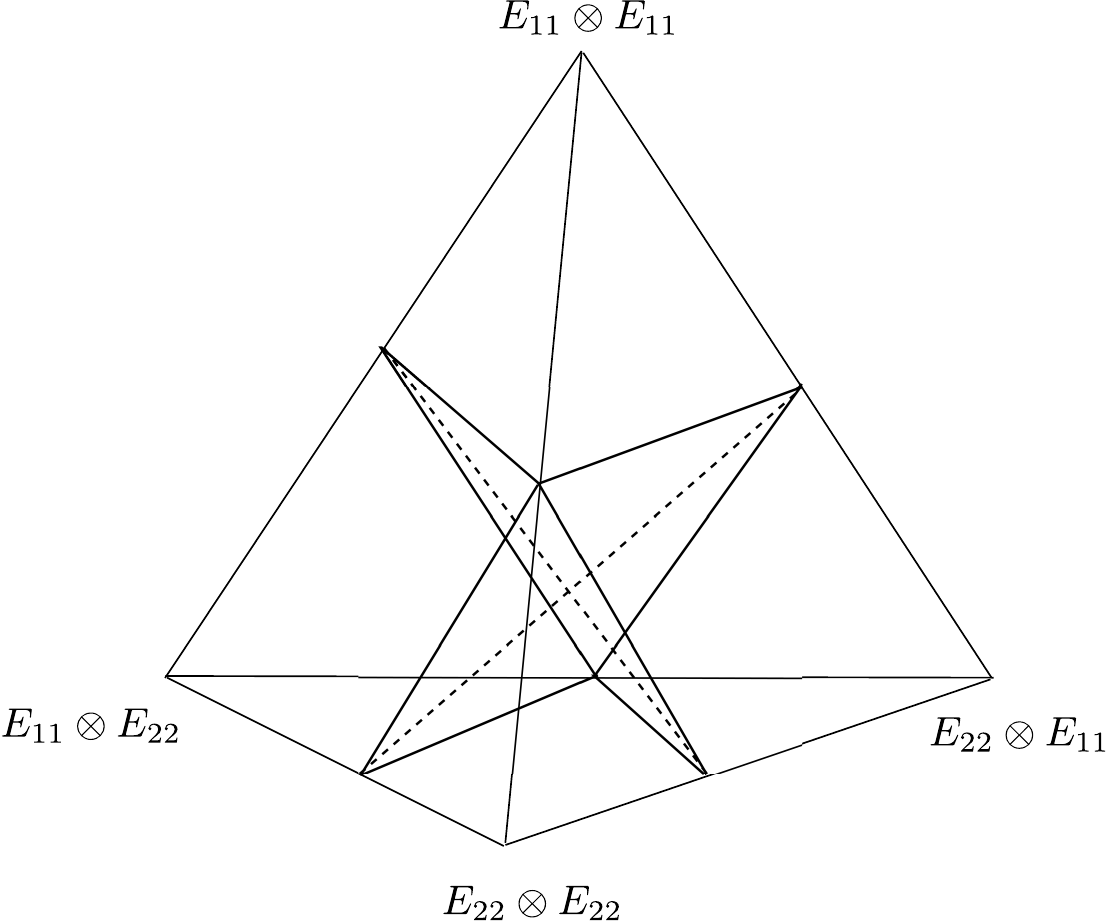}
\par\end{centering}

\caption{Degrees of degeneracy of $\left.\omega\right|_{K.\rho}$ for orbits
through $CC$ states of two qbits. The thick dashed line: $\mathrm{D}(K.\rho)=4$,
the surfaces between thick solid lines: $\mathrm{D}(K.\rho)=2$, the dotted lines and elsewhere:
$\mathrm{D}(K.\rho)=0$.} %Taken from \cite{OSS14}.}
\end{figure}

\section{Summary and outlook}

One of the basic problems in the theory of quantum correlations is the classification of states with respect to local operations performed independently on subsystems of a given system. In this review we  presented a number of results for (1) local unitary (LU) operations, and (2) SLOCC - Stochastic Local Operations with Classical Communication.  Mathematically, these operations are described by the action of some compact group $K\subset  SU(\mathcal{H})$ in case (1) and its complexification $G = K^\mathbb{C}$ in case (2). The space of pure states (after neglecting 
the global phase) is the projective space $\mathbb{P}(\mathcal{H})$, and 
for mixed states, the space of isospectral density matrices is an adjoint 
orbit of the unitary group $SU(\mathcal{H})$. In both cases, these spaces have a natural geometric (K\"{a}hler) structure, and therefore in particular they are symplectic manifolds. Since the action of a compact group on $M$ preserves the symplectic structure there exists the momentum map. In the cases considered here, the momentum map  assigns to a state of $L$ particles its reduced one-particle density matrices, and therefore is directly related to partial traces over $L-1$ particles. This observation opened new possibilities for the analysis employing tools and methods of symplectic and algebraic geometry. 

We showed how such an approach can be applied to analyze/solve the following problems.

\begin{itemize}
	\item When is information contained in one-particle reduced density matrices sufficient to solve the problem of LU-equivalence?
	\item What are obstacles in the case when this information is insufficient, i e., when the LU-equivalence can not be established upon examining spectra of reduced matrices?
	\item  How many additional invariant polynomials (except those directly derived from one-qubit density matrices) are needed to solve the LU-equivalence problem? How does this number depend on the spectra of reduced matrices?
	\item  How to classify states under SLOCC operations? Such a classification should be effective, use straightforwardly calculated quantities and yield a finite number of generalized SLOCC classes.
	\item How to characterize geometrically mixed states with zero discord, more specifically $CC$ and $CQ$ states? 
\end{itemize}

There is a number of open problems that can be explored using the mathematical methods presented here. First, the number of parameters needed, in addition to the single-particle information,  to decide LU-equivalence of quantum states has been presented only for pure qubit states. It would be interesting to extend these results to mixed states and to the cases when local dimensions have dimension smaller greater than two. However, to realise to this aim one would have to first compute Kirwan polytopes for the scenarios in question. Moreover, it would be interesting to interpret physically the action of the complexified group $G$ on the manifold of isospectral density matrices. Another context, where similar geometrical methods can be useful are the scenarios involving entanglement manipulation of delocalised particles \cite{Markus1,Markus2}.

The proposed methods are, in principle applicable to distinguishable as well as indistinguishable particles but concrete results for the latter are scarce. This is another direction of possible further studies. 

Another direction of future research may concentrate on looking for other ways to identify critical orbits important for SLOCC equivalence, for example by employing other approach to stratification described in Section~\ref{sec:GITa} \cite{Hesselink79}. It should allow moving a state to the appropriate critical orbit case a one parameter SLOCC subgroup and thus have a clear operational meaning since one could decide to which class a state belongs by acting on it by a one parameter
family of SLOCC (hence physical) operations.  

\section*{Appendix. Group actions on symplectic manifolds. Momentum maps}

A symplectic manifold is an abstract generalization of a phase-space in classical mechanics. In the following we will invoke notions known in classical mechanics to illustrate some features stemming from this generalizations. Mathematically, a symplectic manifold is a smooth manifold $M$ equipped with a closed, $d\omega=0$, and non-degenerate differential two-form $\omega$. The nondenegeracy means that if $X$ is a tangent field to $M$ such that at each $p\in M$ and each tangent vector $Y$ at $p$ we have $\omega(X,Y)=0$ then $X$ must vanish everywhere on $M$. It implies that a symplectic manifold is always even-dimensional. 

A \textit{K\"ahler manifold} is a complex symplectic manifold equipped with additional Riemannian structure $g$ (a metric, i.e. positive-definite symmetric two-form), such that all three structures (complex, symplectic and Riemannian) are compatible \textit{vis.}, $g(u,v)=\omega(u,iv)$, or, equivalently, $\omega(u,v)=g(iu,v)$. It also means that $h:=g+i\omega$ is a Hermitian metric i.e., $h(iu,iv)=h(u,v)$

A symplectic manifold $(M, \omega)$ is a natural
geometric structure for classical Hamiltonian mechanics. A symplectic
manifold is a classical mechanical phase-space equipped with the
structure of Poisson brackets. They are defined in terms of $\omega$ in
the following way. For an (appropriately smooth) function $F$ on the
phase space we define a tangent vector field $X_F$ \textit{via}
\begin{equation}\label{eq:hamvfield}
dF=\omega(X_F,\cdot),
\end{equation}
i.e.\ the action of the two form $\omega$ on the
vector field $X_F$ gives a one form (as it should) equal to the
differential of $F$. Then we set for the Poisson bracket of two
functions $F$ and $H$,
\begin{equation}\label{eq:poissonbr}
\{F,H\}=\omega(X_F,X_H)
\end{equation}
The dynamics of a system is determined by a Hamilton function $H$ \textit{via} canonical (Hamilton) equations of motion 
\begin{equation}\label{eq:ham}
\frac{d}{dt}F=\{F,H\}
\end{equation} 
for an arbitrary phase-space function $F$. Using now the definitions of the fields $X_H$, $X_F$ and the Poisson brackets we have.
\begin{equation*}
\frac{d}{dt}F=\{F,H\}=\omega(X_F,X_H)=dF(X_H)=X_H(F),
\end{equation*}
hence the vector field $X_H$ determines at each point $p\in M$ the direction in which this point moves under the dynamics generated by the Hamilton function $H$.
  
The momentum map appears always when a Lie group $K$ acts on a symplectic
manifold preserving the symplectic structure. In classical mechanics this corresponds to a situations when we have a group of canonical (i.e. preserving the Poison brackets) symmetries. In this paper $K$ is always a connected compact semi-simple matrix Lie group, in fact a subgroup of a unitary group.  Let us denote the action of $K$ on $M$ by
$x\mapsto\Phi_g(x)$, $x\in M$, $g\in K$. For each element of the Lie
algebra $\xi\in\mathfrak{k} = Lie(K)$ of the group $K$ we define the
\textit{fundamental vector field} on $M$,
\begin{equation}\label{eq:fundvfield}
\hat\xi(x)=\frac{d}{dt}\bigg|_{t=0} \Phi_{\exp t\xi}(x)\ .
\end{equation} 	
Invoking again the classical mechanical origin of the presented concepts we may call $\xi$ a generator of the one dimensional subgroup of symmetries. The vector field $\hat{\xi}$ at $p\in M$ points in the direction in which the phase-space point $p$ moves under the action of this one-parameter subgroup.        

Under additional conditions\footnote{$K$ is semi-simple and $M$ has the trivial first de Rham cohomology group.} fulfilled in all cases considered here, there is a well defined function $\mu_{\xi} $, such that $d\mu_{\xi}
=\omega(\hat{\xi},\cdot)$. Upon referring to the definitions above, we see thus that $\mu_\xi$ plays a role of a Hamilton function for the `motion' of phase-space points under the one-parameter subgroup in question, where the parameter plays a role of the time. In classical mechanics $\mu_xi$ would be called a generating function for the one parameter group.

The functions $\mu_{\xi} $ can be chosen to
be linear in $\xi\in\mathfrak{k}$ and thus define the unique map
$\mu:M\rightarrow\mathfrak{k}^{\ast}$, where $\mathfrak{k}^{\ast}$ is
the dual vector space of $\mathfrak{k}$, i.e.\ the space of linear forms on the linear space $\mathfrak{k}$. Hence, $\mu(x)$ is defined by
$\langle\mu(x),\xi\rangle=\mu_{\xi}(x)$, where $\langle\alpha,\zeta\rangle$ denotes the action of a form $\alpha$ on a vector $\zeta$. The function $\mu:M\rightarrow\mathfrak{k}^{\ast}$ is is called the {\it momentum map}.

The group $K$ acts also on its Lie algebra $\mathfrak{k}$ by the
adjoint action ${\Ad}_{g}\xi=g\xi g^{-1}$. It has its natural dual action on the dual space $\mathfrak{k}^{\ast}$, the coadjoint action
\begin{equation}\label{eq:Adast}
\langle\mathrm{Ad}^\ast_g\alpha,\xi\rangle=\langle\alpha,\mathrm{Ad}_{g^{-1}}
\xi\rangle=\langle\alpha,g^{-1}\xi g\rangle, \quad g\in K, \quad
\xi\in\mathfrak{k}, \quad \alpha\in\mathfrak{k}^\ast.
\end{equation}

The set of points obtained by the action of a group on a manifold on a particular point $p$ we call the orbit of the group (thought the point $p$). In particular the set $\mathcal{O}_p=\SET{q=\Phi_g(p)}{g\in K}\subset M$ is the orbit through a point $p$ of the action of $K$ on $M$, and $\Omega=\SET{\beta=\mathrm{Ad}^\ast_g\alpha}{g\in K}\subset\mathfrak{k}^\ast$ - a coadjoint orbit of $K$ through $\alpha\in\mathfrak{k}^\ast$. 

Each coadjoint orbit is equipped with a natural symplectic structure given by the so-called Kirillov-Kostant-Souriau form. To define it let us first construct the fundamental vector field for the coadjoint action of the one-parameter subgroup of $K$ generated by a Lie-algebra element $\xi$ at some point $\alpha\in\mathfrak{k}$,
\begin{gather}\label{eq:fvfcAd}
\tilde{\xi}(\alpha)=\frac{d}{dt}\bigg|_{t=0}\mathrm{Ad}^\ast_{exp(tX)}\alpha.
\end{gather}
Now we define the Kirillov-Kostant-Souriau form at each point $\alpha$ on a coadjoint orbit by
\begin{equation}\label{symformcoa}
\omega\left(\tilde{\xi}(\alpha),\tilde{\zeta}(\alpha)\right)=\left\langle\alpha,[\xi,\zeta]\right\rangle,
\end{equation}    
where$[\cdot,\cdot]$ is the Lie bracket in the Lie algebra $\mathfrak{k}$. For a semi-simple $K$ the momentum map is equivariant, i.e.
$\mu\left(\Phi_{g}(x)\right)={\Ad}_{g}^{\ast}\mu(x)$ for any $x\in M$
and $g\in K$. Orbits of $K$-action on $M$ are therefore mapped by
$\mu$ onto orbits of the coadjoint action in $\mathfrak{k}^{\ast}$.

Coadjoint and adjoint orbits (i.e.\ the orbits of the adjoint action of $K$ on its Lie algebra $\mathfrak{k}$) can be identified in all cases we consider in the paper\footnote{In fact, the similar identification can be carried out for Lie algebras of {\it arbitrary} compact Lie group.} (i.e.\ for $K$ compact as a subgroup of a unitary group) using a non-degenerate scalar product on $\mathfrak{k}$ defined by $(\xi,\zeta)=-\tr\xi\zeta$, i.e, identifying a linear form $\alpha\in\mathfrak{k}$ with the vector $\xi\in\mathfrak{k}$ such that  $(\xi,\zeta)=\left\langle\alpha,\zeta\right\rangle$ for all $\zeta\in\mathfrak{k}$. We use this identification throughout the paper and therefore we treat the moment map $\mu$ as a map from $M$ to $\mathfrak{k}$ rather than $\mathfrak{k}^\ast$.

\section*{Acknowledgments}

The authors gratefully acknowledge the support of the ERC Grant QOLAPS. AS acknowledges the support from the Marie Curie International Outgoing Fellowship. TM is also supported by Polish Ministry of Science and Higher Education “Diamentowy Grant” no. DI2013 016543.
M.O  acknowledge support from the European Research Council (CoG QITBOX), Spanish MINECO (QIBEQI FIS2016-80773-P,  and Severo Ochoa Grant No. SEV-2015-0522), Fundaci\'o Privada Cellex, and Generalitat de Catalunya (Grant No. SGR 874, 875, and CERCA Programme).

\end{document}